\documentclass{vgtc}                          % final (conference style)
\graphicspath{{figures/}{pictures/}{images/}{./}} % where to search for the images

\usepackage{times}                     % we use Times as the main font
\usepackage{tabu}                      % only used for the table example
\usepackage{booktabs}                  % only used for the table example
\usepackage{lipsum}                    % used to generate placeholder text
\usepackage{mwe}                       % used to generate placeholder figures
\usepackage{mathptmx}                  % use matching math font
\usepackage{graphicx}
\usepackage{subcaption}
\usepackage{amsmath}
\onlineid{1141}

\vgtccategory{Research}

\vgtcinsertpkg

\title{CyberSelf: Embodied Self-Distancing for Emotional Support in Virtual Reality}

\author{
Bing Li\thanks{bing.li@mail.sdu.edu.cn}\\
\scriptsize Shandong University
\and
Yan Hu\thanks{yanhu1985@gmail.com}\\
\scriptsize School of computing
Blekinge Institute of Technology
Karlskrona, Sweden
\and
Tinghui Li\thanks{tili4998@uni.sydney.edu.au}\\
\scriptsize University of Sydney
\and
Yinuo Zhang\thanks{vermonth1216@gmail.com}\\
\scriptsize Shandong University
\and
Wen Ma\thanks{mawen@sdu.edu.cn}\\
\scriptsize Shandong University
\and
Yuanfeng Zhou\thanks{yfzhou@sdu.edu.cn}\\
\scriptsize Shandong University
\and
Yiran Shen\thanks{yiran.shen@sdu.edu.cn }\\
\scriptsize Shandong University
}

\newcommand*\systemname{{\bf CyberSelf}}

\newcommand*\mirror{{\sl Doppelgänger}}

\teaser{
  \centering
  \includegraphics[width=\linewidth]{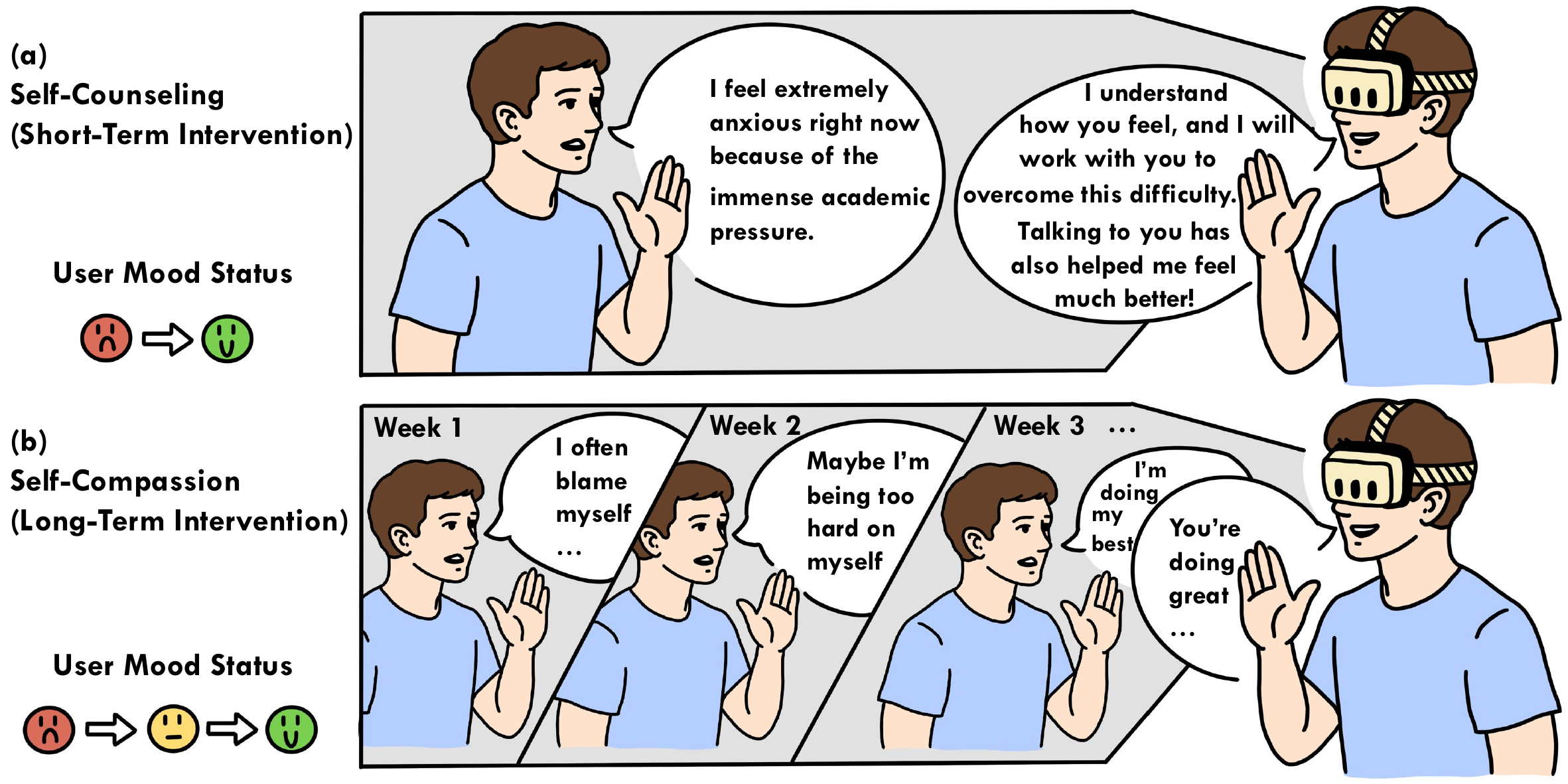}
  \caption{Conceptual overview of \systemname{}, an embodied self-distancing system for emotional support in VR.
(a) In the self-counseling scenario, users engage in short-term dialogue with a self-resembling avatar (\mirror{}) to reflect on immediate distress and support mood relief. (b) In the self-compassion scenario, users interact with \mirror{} repeatedly over time, gradually reframing self-critical thoughts into self-kindness and building longer-term emotional resilience.}
  \label{fig:teaser}
}

\abstract{
Self-distancing is an effective emotion regulation strategy; however, it may fail during personal crises due to its cognitive demands. Virtual Reality (VR) provides a novel approach to externalizing psychological distance by enabling embodied self-representation. 
In this paper, we present \systemname{}, a VR system for emotional support that integrates a visually self-resembling avatar, a cloned self-voice, and Large Language Model (LLM)-driven real-time dialogue. The system enables users to engage in multi-turn conversations with their self-representations in immersive VR, enabling embodied self-distancing while maintaining a strong sense of self-relevance. 
We evaluated \systemname{} in a short-term study that compares three levels of self-representation  richness (Text, Text+Voice, and Text+Voice+Appearance). The results demonstrated robust pre-post improvements across affective and coping measures, specifically increased valence, arousal, hope, and resilience, as well as reduced anxiety and simulator sickness. Richer representations increased conversational engagement, and full embodiment produced the strongest physiological indicators of emotional regulation. A subsequent four-week long-term study demonstrated that these benefits are both sustainable and cumulative. Additionally, users rated the reconstructed avatar and the cloned voice as highly recognizable and acceptable. Collectively, these findings suggest that embodied, self-resembling conversational agents provide a viable mechanism for externalizing self-distancing and supporting emotional regulation in VR. 
} % end of abstract


\begin{document}
\firstsection{Introduction}

\maketitle

%% \section{Introduction} %for journal use above \firstsection{..} instead
Although people often provide effective emotional support to others by adopting an objective perspective, they frequently struggle to apply the same clarity to their own circumstances. When facing with personal difficulties, people often find it difficult to view their situations objectively, a phenomenon associated with the \emph{fundamental attribution error} in self-perception~\cite{FundamentalAttributionError1985}.  Empirical studies indicate that shifting from a self-immersed to a self-distanced perspective can reduce emotional reactivity~\cite{KrossAyduk2011,AydukKross2010}. In particular, referring to oneself in the third person can enhance the ability to reason about personal challenges, a phenomenon known as \emph{Solomon's Paradox}~\cite{GrossmannKross2014}.

Virtual Reality (VR) offers a promising medium for supporting self-distanced reflection by providing immersive and embodied experiences that are difficult to achieve through internal cognitive strategies alone. By enabling interactions with self-resembling avatars~\cite{Guo2024Doppelganger,wang2024survey3dhumanavatar,kolotouros2024instant}, VR enables users to explore alternative perspectives of themselves in a vivid and interactive context. This approach to embodied self-representation allows individuals to engage with personally relevant situations from a visually and spatially distinct perspective, while maintaining a strong connection to their own identity. Previous work on the "Doppelg\"anger Effect"~\cite{Kleinlogel2021Doppelganger} further suggests that observing a virtual representation closely resembling oneself can improve self-relevance and enhance engagement, resulting in meaningful cognitive and behavioral responses~\cite{Guo2024Doppelganger}. Collectively, these attributes establish VR as a compelling platform for supporting vivid, embodied, and personally meaningful forms of self-distanced reflection.

Many existing approaches remain limited in their interactive scope. For example, emerging research on Artificial Intelligence(AI)-generated self-voices has demonstrated promising effects~\cite{Huang2023EmotionalSelfVoice}. However, such audio-only or static visual interventions often lack the multimodal integration and conversational depth required for sustained, meaningful interaction. Without a dynamic and visually embodied presence, these experiences tend to resemble one-sided reflection rather than genuine dialogue. To fully realize the potential of this medium, there is a need for systems that combine high-fidelity visual embodiment with autonomous, real-time interactivity that can support complex, multi-turn emotional conversations. Addressing this gap would enable VR systems to move beyond simple visualization, fostering more responsive, engaging, and emotionally resonant forms of interaction.

In this paper, we present \systemname{}, a novel system that integrates high-fidelity visual self-cloning with AI-driven emotional voice synthesis to create a \mirror{} within a fully immersive VR environment. Unlike previous approaches that rely on pre-scripted interactions or audio-only, one-shot responses~\cite{Huang2023EmotionalSelfVoice}, \systemname{} utilizes a large language model (LLM) to generate autonomous, context-aware, and emotionally attuned dialogue in real time. By combining the psychological salience of the \mirror{} effect with VR-based embodiment and the generative capabilities of modern AI, \systemname{} enables dynamic, multi-turn conversations with an embodied ``idealized self'', facilitating a more engaging and meaningful self-distancing experience.
We evaluated \systemname{} through both short-term and long-term studies conducted in two self-distancing scenarios, \emph{self-counseling} and \emph{self-compassion}. In both settings, participants reported enhanced emotional support, with stronger effects observed in richer forms of self-representation and evidence of sustained benefits with repeated use.

Overall, \systemname{} provides an autonomous and immersive self-distancing framework that translates established psychological theory into an interactive and deployable experience. The contributions of this paper are threefold:

\begin{itemize}
    \item We design and implement \systemname{}, a novel VR system featuring an embodied \mirror{}, which closely replicates the user’s appearance and voice. This system is designed to facilitate self-distanced reflection for individuals experiencing emotional challenges. To the best of our knowledge, this study is the first to systematically investigate self-distancing through an embodied \mirror{} in VR.
    
    \item We synthesize high-fidelity visual and vocal self-representations empowered by generative AI and LLMs, which enables seamless, real-time, multi-turn conversations between users and their embodied \mirror{}. This multimodal integration enhances immersion and engagement in self-distanced interaction.
    
    \item We conduct both short-term and long-term user studies in two self-distancing scenarios to validate the system. The results demonstrate that \systemname{} effectively supports emotional support, with richer self-representation (\textbf{Text+Voice+Appearance}) leading to stronger perceived support and engagement compared to \textbf{Text}-only or \textbf{Text+Voice} baselines. Furthermore, repeated use results in sustained benefits in self-compassion over time.
\end{itemize}

\section{Related Work}
\label{sec:related}

% Our work is inspired by prior research on embodied self-distancing, self-resembling avatars in VR, and AI-mediated emotional support through conversational agents. We review these areas and position \systemname{} relative to existing approaches.

% \subsection{Self-Distancing Interventions for Emotion Regulation}

% Empirical research in psychology has shown that self-distanced reflection facilitates emotional recovery by enabling individuals to analyze negative experiences without reactivating intense affect. O. Ayduk and E. Kross~\cite{AydukKross2010} found that individuals who spontaneously adopted a distanced perspective when reflecting on negative events exhibited lower emotional distress and reduced rumination over time, highlighting the regulatory benefits of perspective change. Related studies~\cite{KrossAyduk2011} have further demonstrated that linguistic forms of self-distancing, such as using one’s own name or third-person pronouns, can reduce emotional reactivity during stressful or threatening situations.

% Beyond affective regulation, self-distancing has also been linked to higher-order reasoning and judgment. Grossmann and Kross~\cite{GrossmannKross2014} introduced the concept of \emph{Solomon’s Paradox}, showing that people tend to reason more wisely about others’ problems than about their own, and that adopting a distanced perspective can partially alleviate this bias when reasoning about personal dilemmas. Informed by these theoretical foundations, we adopt self-distancing interventions as the core design principle for \systemname{}. 

\subsection{Self-Distancing Interventions for Emotion Regulation}

Empirical research in psychology shows that self-distanced reflection facilitates emotional recovery by enabling individuals to analyze negative experiences without reactivating intense affect. Ayduk and Kross~\cite{AydukKross2010} found that individuals who spontaneously adopted a distanced perspective when reflecting on negative events exhibited lower emotional distress and reduced rumination over time, highlighting the regulatory benefits of perspective change. Related studies~\cite{KrossAyduk2011} further demonstrate that linguistic forms of self-distancing, such as using one’s own name or third-person pronouns, can reduce emotional reactivity during stressful or threatening situations.
Beyond affective regulation, self-distancing has also been linked to higher-order reasoning and judgment. Grossmann and Kross~\cite{GrossmannKross2014} introduced the concept of \emph{Solomon’s Paradox}, showing that people tend to reason more wisely about others’ problems than their own, and that adopting a distanced perspective can partially alleviate this bias when reasoning about personal dilemmas. Informed by these findings, we adopt self-distancing interventions as the core design principle of \systemname{}.

\subsection{Self-Avatar in Virtual Reality}

Osimo et al.~\cite{Osimo2015SigmundFreud} introduced a virtual body-ownership paradigm in which users alternated embodiment between their own virtual body and a counselor figure modeled after Sigmund Freud. Conversing with their embodied double enabled externalized self-dialogue and reduced negative affect.
Subsequent work explored self-resembling avatars in collaborative and training contexts. Kleinlogel et al.~\cite{Kleinlogel2021Doppelganger} proposed a doppelg\"anger-based training paradigm in which users imitated their virtual selves to accelerate interpersonal skill learning. Guo et al.~\cite{Guo2024Doppelganger} similarly found that collaboration with a self-similar virtual character improved task performance and perceived cooperation.
Research has also examined how personalization and synchrony influence embodiment and user experience. Jung et al.~\cite{JungEtAl2022PersonalizedAvatarMotion} showed that self-resembling avatars and motion synchrony enhance embodiment, presence, and enjoyment. Increasing avatar self-similarity across appearance, voice, and identity cues further strengthens embodiment and social presence, though effects on immersion are mixed~\cite{KimParkLee2023SelfSimilarSocialVR}. Kleinlogel et al.~\cite{KleinlogelEtAl2024MeetSelfGender} additionally reported that encountering one’s virtual self can evoke ambivalent reactions moderated by gender.
Beyond appearance, prior work demonstrates that user--avatar movement coupling can be manipulated to influence embodiment. Li et al.~\cite{LiEtAl2022MovementInconsistency} showed that subtle movement inconsistencies alter body ownership without breaking immersion, while Ahuja et al.~\cite{AhujaEtAl2021CoolMoves} introduced motion accentuation techniques that enhance expressiveness while preserving agency and embodiment.

% Prior work has primarily examined self-avatars in the context of learning, collaboration, embodiment, and general social interaction. While some studies report emotional effects, they do not explicitly frame self-avatar interaction as a self-distancing intervention for emotion regulation, nor do they leverage autonomous, multi-turn dialogue to support reflective self-counseling. Therefore, in this study, we explicitly integrates self-resembling avatars with the psychological principle of self-distancing and an AI-driven conversational agent to enhance the intervention. 

\subsection{Generative AI for Emotional Support}

A substantial body of research has explored conversational agents for emotional and mental health support. Early work on relational agents emphasized the importance of empathy, rapport, and continuity for establishing therapeutic alliances in automated systems~\cite{Bickmore2010RelationalAgentsPsychiatry}. Virtual human systems such as \emph{SimSensei} further showed that face-to-face interaction and nonverbal behaviors can facilitate disclosure and increase user comfort and engagement~\cite{DeVaultEtAl2014SimSensei}.
Text-based agents have also been widely studied due to their scalability. Systems such as Reflection Companion support reflection through adaptive dialogues~\cite{KocielnikEtAl2018ReflectionCompanion}, while reviews of social-emotional learning agents highlight limitations including shallow interaction and limited feedback~\cite{FuEtAl2022SELAgents}. Meta-analyses indicate that mental health chatbots can produce modest improvements in depression, anxiety, and psychological distress, though outcomes vary across studies~\cite{AbdAlrazaqEtAl2020ChatbotsMeta,LiEtAl2023AICAsMeta}.
The emergence of LLMs has renewed interest in generative emotional support. Recent surveys report improvements in fluency, coherence, and perceived empathy compared to rule-based systems~\cite{JinEtAl2025LLMMentalHealthScoping}. Newer systems explore personalized and multimodal approaches, including AI- and AR-enhanced storytelling for social-emotional learning~\cite{LyuEtAl2024EMooly}. Some work has begun incorporating self-relevant cues, such as AI-generated emotional self-voice to promote reflection and behavior change~\cite{Huang2023EmotionalSelfVoice}; however, these approaches remain limited to audio-based interaction and lack embodied visual self-representation.

% Therefore, to enhance the sense of self-representation, we integrate generative AI-driven, real-time dialogue with an embodied, self-resembling avatar in immersive VR. Rather than positioning the AI as an external counselor or generic conversational partner, our system enables sustained, multi-turn interaction with an embodied representation of the self. By combining LLM-based conversational intelligence with visual and vocal self-similarity, \systemname{} explicitly leverages embodied self-distancing as a mechanism for emotional regulation, an aspect that prior generative AI--based emotional support systems have failed to incorporate.

\section{System Design}

\begin{figure}[htb]
  \centering
  \includegraphics[width=\linewidth]{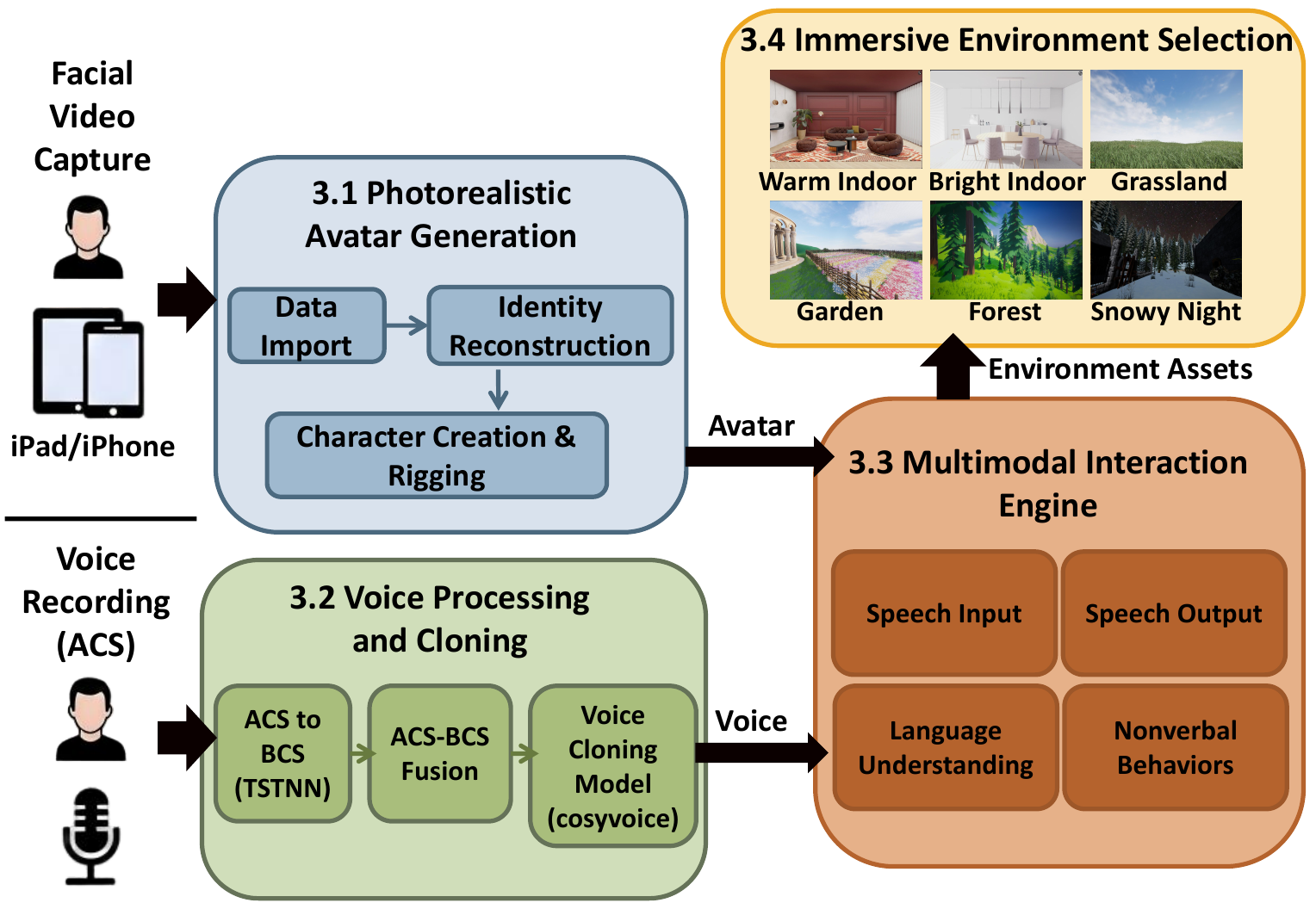}  % 图片的路径
  \caption{the overall system architecture of \systemname{}.}
  \label{fig:system}
\end{figure}

% We developed \systemname{}, a high-fidelity VR self-distancing platform that delivers AI-driven emotional support in an immersive environment. As shown in Fig.~\ref{fig:system}, \systemname{} comprises four primary modules: (1) \textbf{Photorealistic Avatar Generation} module utilizes Unreal Engine~5 (UE5)~\cite{UnrealEngine5} to reconstruct a user-resembling avatar, referred to as the \mirror{}, from a short facial video clip; (2) \textbf{Voice Processing and Cloning} module combines audio pre-processing with a neural voice-cloning pipeline to synthesize speech in the user’s own voice; (3) \textbf{Multimodal Interaction Engine} module integrates speech input/output, nonverbal behaviors, and an LLM to enable natural, human-like, multi-turn conversations; and (4) \textbf{Immersive Environment Selection} module provides carefully designed VR settings to promote comfort, presence, and overall user experience. 
% The system runs on a Pico~4~Pro headset~\cite{PICO4}.

We present \systemname{}, a VR self-distancing system for AI-driven emotional support. As shown in Fig.~\ref{fig:system}, it includes four components:
(1) \textbf{Avatar Generation}, which uses Unreal Engine~5 (UE5)~\cite{UnrealEngine5} to reconstruct a user-resembling \mirror{} from a short facial video;
(2) \textbf{Voice Processing and Cloning}, which synthesizes speech in the user’s voice;
(3) \textbf{Multimodal Interaction}, which integrates speech I/O, nonverbal behaviors, and an LLM for real-time multi-turn dialogue; and
(4) \textbf{Environment Selection}, which provides immersive VR settings.

\subsection{Photorealistic Avatar Generation Module}
\label{sec:Photorealistic Avatar Generation}

\begin{figure}[h]
  \centering
  \includegraphics[width=\linewidth]{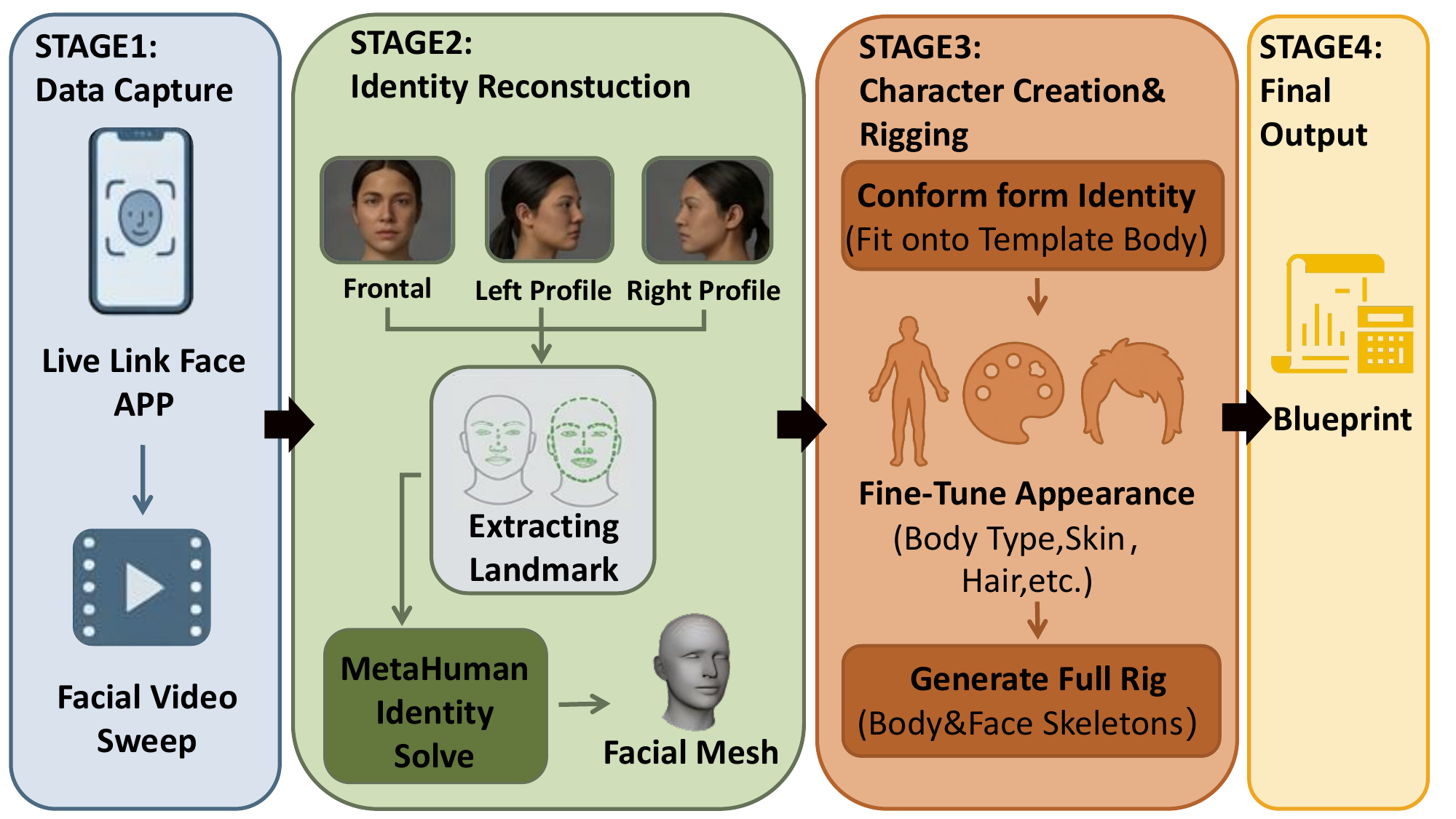}  % 图片的路径
  \caption{The overall pipeline of Photorealistic Avatar Generation module}
  \label{fig:avatar_generation_pipeline}
\end{figure}
% To improve the self-relevance during self-distancing, it is essential to minimize the perceptual mismatch between the physical user and their cyber \mirror s. We therefore build our avatar pipeline on the MetaHuman framework, with an emphasis on high-fidelity facial reconstruction, which serves as the primary cue for identity recognition. This  approach facilitates users' perception of the avatar as representing themselves. To maintain overall realism while avoiding introducing unnecessary modeling effort or variability, the personalized face is integrated with standard MetaHuman body templates for the remainder of the avatar.

To maintain strong self-relevance during self-distancing, we minimize perceptual mismatch between users and their cyber \mirror{}. Our pipeline is built on the MetaHuman framework, emphasizing high-fidelity facial reconstruction, which is the primary cue for identity recognition. To maintain overall realism while avoiding introducing unnecessary modeling effort,  the personalized face is integrated with standard MetaHuman body templates.

As shown in Fig.~\ref{fig:avatar_generation_pipeline}, we capture a short facial sweep (frontal to profiles) using the Live Link Face app~\cite{LiveLinkFaceApp}. In UE5, the facial landmarks from three views (frontal, left, right) are extracted to solve user-specific facial geometry. The resulting MetaHuman-compatible facial mesh is then applied to a full character using Conform from Identity. Finally, appearance attributes are refined and a full control rig is generated, producing a deployable avatar blueprint for VR scenes. 

\subsection{Voice Processing and Cloning Module}
\label{sec:Voice Processing and Cloning}

Another key aspect of the \mirror{} is the \emph{own-voice effect}. Prior work leveraged voice cloning to generate an ``ideal-self'' speaking in the user's voice~\cite{Huang2023EmotionalSelfVoice}. However, the perceived self-voice combines both air-conducted speech (ACS) and bone-conducted speech (BCS), whereas microphones capture primarily ACS, which can lead to voice confrontation. To address this, our \textbf{Voice Processing and Cloning} pipeline reconstructs a perceptually familiar own-voice signal by estimating the missing BCS component before voice cloning.

The participants recorded four sentences from the RASC863 corpus~\cite{Wang2022ABCS} using a smartphone recorder (the specific sentences used can be found in the supplementary materials). To estimate BCS, we adopt the Two-Stage Transformer Neural Network (TSTNN)~\cite{kim2025tapsthroatacousticpaired}, trained on the Throat and Acoustic Paired Speech (TAPS) dataset~\cite{kim2025tapsthroatacousticpaired}. Given a recorded ACS signal, the model predicts the corresponding BCS component $S_{\text{BC}_{\text{pred}}}(t)$.

To suppress noise artifacts, we apply DeepFilterNet2~\cite{schröter2022deepfilternet2realtimespeechenhancement}, producing $\tilde{S}_{\text{BC}_{\text{pred}}}(t)$. Following~\cite{BCSacs2021}, the perceived self-voice is modeled as a weighted combination of ACS and BCS:
\begin{equation}
S_{\text{self}}(t) = \gamma S_{\text{AC}}(t) + (1-\gamma)\tilde{S}_{\text{BC}_{\text{pred}}}(t)
\end{equation}
where $\gamma$ balances air- and bone-conducted components. We set $\gamma=0.35$ following~\cite{BCSacs2021}. The fused signal serves as the canonical voice representation.

Finally, neural voice cloning is performed using CosyVoice~\cite{du2025cosyvoice3inthewildspeech}, which supports zero-shot speaker adaptation. A short enrollment clip (10--15\,s) is used to extract a speaker embedding, enabling the \mirror{} to synthesize speech in the user's voice.

\subsection{Multimodal Interaction Engine Module}
\label{sec:multi-interaction}
% To enhance user engagement in \systemname{}’s immersive environment, an LLM-driven multimodal interaction engine is designed to coordinate speech input, language understanding, speech output, and nonverbal behaviors. This engine enables real-time, coherent, and emotionally resonant dialogue. It prioritizes conversational continuity and low-latency responsiveness, both of which are essential for sustaining immersion and supporting natural interaction in a self-distancing context.

To support natural self-dialogue, \systemname{} employs an LLM-driven multimodal interaction engine that coordinates speech input/output, language understanding, and nonverbal behaviors to enable real-time, coherent conversation.

\begin{figure}[h]
  \centering
  \includegraphics[width=\linewidth]{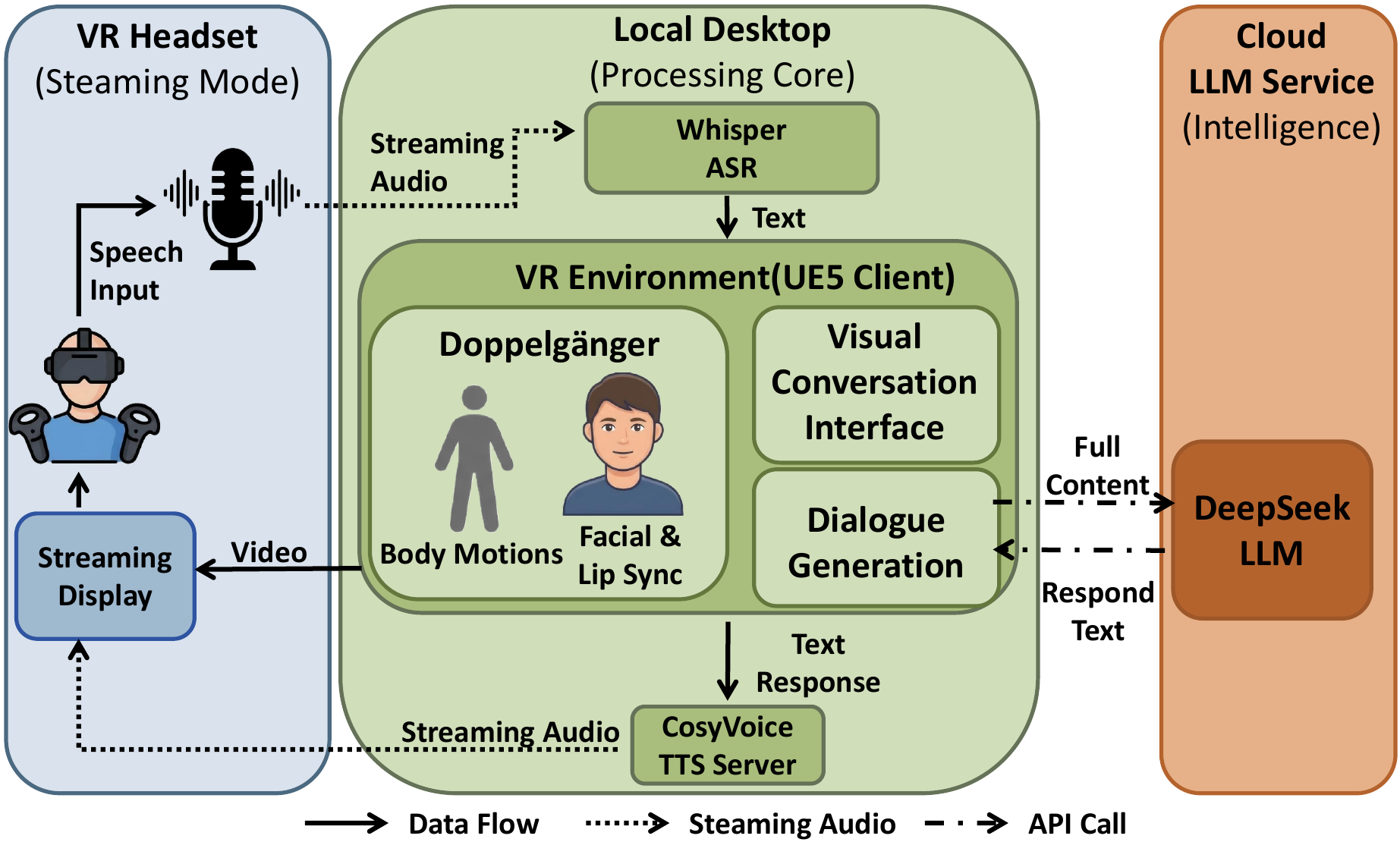}  % 图片的路径
  \caption{Multimodal Interaction Engine Architecture.}
  \label{fig:multimodol interaction}
\end{figure}

% As shown in Fig.~\ref{fig:multimodol interaction}, the multimodal interaction engine consists of components distributed across multiple platforms: a VR headset operating in streaming mode, a local desktop that responsible for most on-device processing (from audio capture to speech-to-text), and a remote cloud service that provides LLM-based dialogue generation. Specifically, user speech is captured by the VR headset’s embedded microphones and streamed to the local desktop for processing. The audio is then transcribed using a locally deployed instance of OpenAI’s Whisper~\cite{OpenAIWhisper}, an automatic speech recognition (ASR) module that converts speech into text. The resulting transcript is subsequently forwarded to the dialogue generation component.

As illustrated in Fig.~\ref{fig:multimodol interaction}, the system distributes processing across three components: a VR headset, a local desktop, and a cloud LLM service. User speech captured by the headset microphones is streamed to the desktop, where it is transcribed using a local deployment of Whisper ASR~\cite{OpenAIWhisper}. The transcript is then sent to the dialogue generator powered by the DeepSeek LLM~\cite{DeepSeekV3}, selected for its strong Chinese language capabilities.

% Since interactions are conducted primarily in Chinese, the dialogue generation core is powered by the DeepSeek LLM~\cite{DeepSeekV3}, for its strong Chinese language understanding and generation capabilities. 
% % which provides the system’s conversational intelligence. 
% % DeepSeek is selected 
% The model is accessed remotely by calling its API service, however, the API is stateless and does not retain conversational context across requests. To support multi-turn coherence and conversational memory, we explicitly manage dialogue history within UE5: each user utterance and system response is appended to an array-based message buffer, and the full conversation context is included in every API call. This design enables \systemname{} to generate contextually consistent and emotionally appropriate responses throughout the dialogue.

Because the LLM API is stateless, dialogue history is maintained in UE5 using a message buffer that stores all user and system utterances. The full conversation context is included in each API request to preserve multi-turn coherence.

% The generated textual response is subsequently converted to speech using CosyVoice~\cite{du2025cosyvoice3inthewildspeech} on the local desktop, which operates as a dedicated text-to-speech (TTS) server, while UE5 functions as the client. Communication between UE5 and CosyVoice is implemented via a WebSocket-based architecture to support low-latency, bidirectional data exchange. Upon system initialization, the WebSocket connection is established and maintained throughout the session. During interaction, UE5 streams textual responses to the CosyVoice server, which returns synthesized audio in a chunked, streaming format. This approach allows audio playback to begin before the full waveform generation completes, allowing users to hear the avatar’s response within approximately 3 seconds after completing an utterance. By contrast, ESV~\cite{Huang2023EmotionalSelfVoice} requires users to wait approximately 30 seconds for responses from a voice-only avatar, which impedes fluid multi-turn conversation.

Generated responses are synthesized into speech using CosyVoice~\cite{du2025cosyvoice3inthewildspeech}, running on the local desktop as a TTS server. UE5 communicates with the server via WebSocket to enable low-latency streaming. Audio is returned in chunks so playback can begin before synthesis completes, allowing responses within 3\,s. This latency is substantially lower than prior voice-only systems such as ESV~\cite{Huang2023EmotionalSelfVoice}, which require 30\,s per response.

To improve dialogue usability, we integrate a visual conversation interface in VR that displays text transcripts of the interaction, allowing users to review prior turns during emotionally reflective conversations. Speech input is controlled through explicit buttons: a green button activates ASR listening and a red button stops capture, providing clear turn-taking and reducing unintended interruptions.

The multimodal engine also coordinates nonverbal behaviors to enhance social presence. Body motions are implemented using ActorCore animation assets retargeted to the MetaHuman skeleton via UE5. Facial expressions (e.g., blinking and eyebrow movements) are driven by predefined animation assets, while lip synchronization is achieved using the Runtime MetaHuman LipSync plugin. These components keep speech, facial articulation, and body motion temporally aligned, producing a coherent and lifelike conversational experience.

\subsection{Immersive Environment Selection Module}
\label{sec:Immersive Environment Selection}

\systemname{} provides six immersive environments that users can select to support emotional regulation and self-distanced reflection. The environments vary along two dimensions: \emph{indoor vs.\ outdoor} and \emph{affective tone}.

Prior environmental psychology studies suggest that enclosed spaces can facilitate self-disclosure and reflective dialogue~\cite{altman1975privacy}. Warm lighting promotes relaxation and reduces perceived stress~\cite{knez1995lighting}, while brighter indoor lighting supports alertness and cognitive engagement~\cite{boyce2014lighting}. Accordingly, we designed two indoor environments (warm-toned and bright-toned).

Guided by Attention Restoration Theory and Stress Recovery Theory, natural environments can promote attentional recovery and emotional restoration~\cite{kaplan_art,ulrich_stress}. We therefore created four outdoor environments (grassland, garden, forest, and snowy night). Participants could preview these environments and select the one that best supported comfort and openness during the self-distancing session.

\section{User Study}

We conducted both short-term and long-term user studies to evaluate \systemname{} in supporting immediate problem-solving and longer-term emotional regulation\footnote{All studies were approved by the local Institutional Review Board.}.

\subsection{Self-distancing Scenarios Description}

% The core strength of \systemname{} is its capacity to facilitate direct, face-to-face dialogue between users and  their \mirror{}, an avatar in a VR environment that closely replicates both their appearance and voice. This high degree of visual and vocal enhances self-identification and embodiment~\cite{KimParkLee2023SelfSimilarSocialVR,Kao2021SelfSimilarVoice}. Building on this capability, the studies were structured around the psychological strategy of  self-distancing to provide emotional support within immersive environments.
The key capability of \systemname{} is enabling face-to-face dialogue between users and their \mirror{}, a VR avatar that closely matches the user’s appearance and voice. This high self-similarity enhances self-identification and embodiment~\cite{KimParkLee2023SelfSimilarSocialVR,Kao2021SelfSimilarVoice}. Building on this property, we design the studies around \emph{self-distancing}, allowing users to interact with their difficulties from an observer perspective.

% Two emotional intervention scenarios, \emph{self-counseling} and \emph{self-compassion}, were developed to examine how embodied self-distancing in VR can support emotional regulation across short- and long-term time scales and various types of emotional challenges. Both scenarios utilize a core interaction paradigm, in which users converse with an embodied \mirror{} from an observer’s perspective, allowing them to externalize and address their own difficulties as if supporting another individual.

We evaluate two emotional intervention scenarios: \emph{self-counseling} and \emph{self-compassion}, targeting short-term problem solving and longer-term emotional challenges, respectively. In both scenarios, users converse with their embodied \mirror{} to externalize personal difficulties and respond to them as if supporting another person.

\subsubsection{\bf Scenario~1: Self-Counseling (Short-Term Intervention)}

The \emph{self-counseling} scenario focuses on short-term problem solving. Participants describe a current difficulty (e.g., academic pressure or decision conflicts), which is used as a prompt for the LLM to guide the \mirror{} in VR. The \mirror{}, resembling the user in appearance and voice, expresses the problem and associated distress. Positioned as an external observer, the user then analyzes the situation and offers advice or coping strategies.

This design is grounded in research showing that self-distanced perspectives improve reasoning about personal problems and reduce the self--other asymmetry described by Solomon’s Paradox~\cite{GrossmannKross2014}. Distancing also helps individuals regulate emotional reactions and evaluate situations more objectively~\cite{KrossAyduk2011}. Because these challenges are typically situational, the \emph{self-counseling} intervention is designed as a single-session experience.

\subsubsection{\bf Scenario~2: Self-Compassion (Long-Term Intervention)}

The \emph{self-compassion} scenario targets deeper emotional challenges related to self-criticism and negative self-evaluation. Self-compassion involves responding to personal suffering with kindness, recognizing it as part of shared human experience, and maintaining balanced awareness of emotions~\cite{Neff2003}. Because such patterns develop over time, we designed this scenario as a long-term intervention in which participants completed weekly VR sessions over one month.

Participants reflect on persistent insecurities or traits they struggle with (e.g., difficulty communicating with others). This information guides the \mirror{} to express vulnerability and self-criticism using the user’s appearance and voice. Confronted with this externalized self, users typically comfort and encourage the \mirror{}, practicing compassion toward themselves from a distanced observer perspective.

\subsection{Conditions}
\label{sec:Conditions}

This study examined how different levels of self-representation in VR influence users’ emotional experience and self-distancing. Three conditions were compared: \textbf{Text}, \textbf{Text+Voice}, and \textbf{Text+Voice+Appearance}. Participants wore a VR head-mounted display in all conditions to ensure a consistent immersive context.

The \textbf{Text} condition served as a baseline, where participants interacted through the conversation interface and received AI responses only in text, without synthesized speech or an avatar.

The \textbf{Text+Voice} condition added voice output using a voice matched to the participant, while interaction remained text-based and no visual self-representation was provided.

The \textbf{Text+Voice+Appearance} condition used the full capability of \systemname{}. Participants engaged in face-to-face dialogue with the \mirror{}, which resembled their appearance and spoke with a cloned voice, integrating textual, auditory, and visual self-representations for a fully embodied self-distancing experience.

\subsection{Measures}
\label{sec:Measures}

To assess participants’ emotional and experiential states, we used validated self-report questionnaires commonly adopted in affective computing, psychology, and HCI research.

\subsubsection{\textbf{Measures for Self-Counseling}}

Immediate emotional responses were measured using \emph{Valence} and \emph{Arousal} ratings on 7-point Likert scales~\cite{Huang2023EmotionalSelfVoice}. Valence reflects emotional positivity or negativity, while arousal captures emotional intensity. \emph{State anxiety} was assessed using the \emph{State--Trait Anxiety Inventory (STAI)}~\cite{Spielberger1983STAI}. \emph{Psychological resilience} was measured using the \emph{Brief Resilience Scale (BRS)}~\cite{Smith2008BRS}, and \emph{motivational outlook} was evaluated using the \emph{State Hope Scale (SHS)}~\cite{Snyder1996SHS}.

To assess potential VR discomfort, we administered the \emph{Simulator Sickness Questionnaire (SSQ)}~\cite{Kennedy1993SSQ}. Participants also completed the \emph{Networked Minds Social Presence scale (NM-SP)}~\cite{Biocca2004NetworkedMindsReliability} to measure perceived realism and interpersonal quality of the interaction.

\subsubsection{\textbf{Measures for Self-Compassion}}

For the long-term \emph{self-compassion} scenario, we used a questionnaire battery focusing on self-compassion and emotion regulation across sessions. Emotional states were again assessed using \emph{Valence} and \emph{Arousal} ratings. \emph{Momentary self-compassion} was measured using the \emph{State Self-Compassion Scale (State SCS)}~\cite{Neff2003SCS}.

Participants’ perceived ability to regulate negative emotions was evaluated using the \emph{Negative Mood Regulation Scale (NMR)}~\cite{Catanzaro1990NMR}. As in the self-counseling study, \emph{SSQ} measured VR discomfort and \emph{NM-SP} assessed perceived social presence.

\subsection{Participants Recruitment}

We recruited 36 participants (23 male, 13 female) through university advertisements and provided compensation proportional to participation time. All participants received a description of the study scenarios and provided informed consent prior to participation.
Among them, 24 participants (12 male, 12 female) completed the short-term \emph{self-counseling} scenario, while 12 participants (11 male, 1 female) completed the long-term \emph{self-compassion} scenario. The short-term sample size aligns with common practice in controlled user studies~\cite{Caine2016}. The long-term study focused on individuals experiencing sustained psychological stress; therefore, participants were primarily recruited from PhD student populations.
Participants were required to be at least 18 years old, fluent in spoken Chinese, and have normal hearing, with no history of neurological disorders or conditions that could interfere with VR use. The final sample consisted of young adults aged 21--29 years (\emph{M}=23.8, \emph{SD}=1.86), a demographic commonly affected by mental health challenges and a key target group for emotional support interventions~\cite{WHOUnicef2024YouthMH}. All participants resided in China at the time of the study.

\subsection{Procedure}
\label{sec:Procedure}

\subsubsection{\bf Onboarding}

\paragraph{\bf Facial and voice data collection.}
Before the dialogue intervention, facial and voice data were collected to reconstruct a personalized avatar and clone each participant’s voice, forming a user-specific \mirror{}. Facial recordings consisted of short video clips (19.921\,s on average, \emph{SD}=6.603\,s), while speech recordings averaged 18.706\,s (\emph{SD}=2.826\,s). The self-avatar and cloned voice were then generated using the pipelines described in Section~\ref{sec:Photorealistic Avatar Generation} and Section~\ref{sec:Voice Processing and Cloning}. Participants subsequently rated the perceived quality of the reconstructed appearance and cloned voice as a manipulation check of self-representation fidelity.

\paragraph{\bf Character prompt preparation.}
To guide the LLM-driven dialogue, participants completed a structured character-prompt template describing: (1) a brief self-description, (2) the situation influencing their emotions, (3) the dialogue goal (e.g., gaining clarity or support), (4) the preferred speaking style of the \mirror{}, and (5) constraints (e.g., no action descriptions, a 300-character response limit, and support for multi-turn dialogue). Participants could also add optional constraints to personalize the interaction.

Prompts were written in the second person (``you'') so the LLM would adopt the role of the \mirror{}. Participants were encouraged to provide truthful and sufficiently detailed information to reduce hallucinated details and maintain conversational coherence. The full template and an example are provided in supplementary materials.

\subsubsection{\textbf{Procedure for Short-Term Self-Counseling}}
\label{sec:Scenario1 Intervention}

In the short-term \emph{self-counseling} intervention, participants first selected a current personal challenge (e.g., a setback or dilemma) when completing the character prompt and were encouraged to focus the dialogue on generating concrete solutions and coping strategies.

% After completing the prompt, participants wore a Polar~H10 chest strap for physiological recording. ECG signals were collected using the \emph{ECGLogger} mobile application via Bluetooth. Participants then donned the VR headset and entered \systemname{}.

After completing the prompt, participants donned the VR headset and entered \systemname{}. Inside the system, participants previewed several virtual environments and selected their preferred setting, which remained constant across all conditions. They then provided demographic information and were instructed to immerse themselves in the negative emotions associated with the described scenario. Once participants reported that these emotions had become sufficiently salient, they completed the pre-intervention questionnaires described in Section~\ref{sec:Measures}. 
% ECG recording began at this stage and continued throughout the intervention until the post-intervention questionnaires were completed.

Participants then engaged in the VR dialogue without a fixed time limit and could end the session once their emotional distress had eased or they had gained sufficient clarity about the problem. Immediately afterward, they completed the same questionnaires to measure pre--post changes in emotional state, along with the \emph{NM-SP} scale to evaluate perceived interaction realism.

The short-term study adopted a \textbf{within-subject design}, where each participant experienced all three conditions. Condition order was counterbalanced using a Latin square scheme~\cite{2023LatinSquare}. Participants took at least a 30-minute break between conditions to minimize carryover effects. The character prompt and selected environment remained constant across conditions, and the same questionnaire battery was administered each time.

% \subsubsection{\textbf{Procedure for Long-Term Self-Compassion}}

% During the long-term \emph{self-compassion} intervention, participants were asked to focus on enduring self-related concerns. For the character prompt, they described their personality traits, perceived shortcomings or aspects of themselves that they found difficult to accept. Participants were instructed to frame the dialogue goal as providing encouragement and comfort to the distressed self.

% The system setup replicated the protocol used in the \emph{self-counseling} intervention. Participants completed questionnaires before and after each session. The long-term assessment battery was specifically designed to evaluate constructs closely associated with self-compassion and sustained emotion regulation (Section~\ref{sec:Measures}). Consistent with the the \emph{self-counseling} intervention, dialogue duration in the \emph{self-compassion} condition was self-paced (i.e., not time-limited). After each conversation, participants again completed the same measures to assess pre--post changes and additionally completed the \emph{NM-SP} to evaluate perceived social presence and interaction realism.

% To explore longitudinal effects in the long-term \emph{self-compassion} study, participants completed one intervention session per week for one month. Only the \textbf{Text+Voice+Appearance} condition was included to isolate the longitudinal effects of \systemname{} and minimize confounding variables associated with repeated exposure to other conditions.

\subsubsection{\textbf{Procedure for Long-Term Self-Compassion}}

In the long-term \emph{self-compassion} intervention, participants focused on enduring self-related concerns. When completing the character prompt, they described personality traits, perceived shortcomings, or aspects of themselves they struggled to accept, and framed the dialogue goal as offering encouragement and comfort to the distressed self.

The system setup followed the same protocol as the \emph{self-counseling} intervention. Participants completed questionnaires before and after each session using the assessment battery described in Section~\ref{sec:Measures}. Dialogue duration was self-paced, and after each conversation participants again completed the measures to assess pre--post changes, along with the \emph{NM-SP} scale to evaluate perceived social presence and interaction realism.

To examine longitudinal effects, participants completed one session per week for one month. Only the \textbf{Text+Voice+Appearance} condition was included to isolate the longitudinal impact of \systemname{} while minimizing confounds from repeated exposure to other conditions.

% \subsection{Analysis Methods}

% A within-subjects design was used for the short-term studies and the same analytical framework was consistently applied to long-term datasets. Our primary analyses used mixed-effects models implemented using \texttt{lme4} package in R~\cite{Bates2015lme4}. Specifically, random-intercept models were employed, allowing each participant to have a unique intercept to account for stable individual differences across repeated measurements (i.e., incorporate individual differences as a random effect). The independent variables included the experimental condition (three levels: \textbf{Text}, \textbf{Text+Voice}, and \textbf{Text+Voice+Appearance}) and time (two levels: pre-experiment and post-experiment).

% To identify the levels of the factors, post-hoc pairwise comparisons were conducted. Specifically, pairwise contrasts between experimental conditions were performed using estimated marginal means implemented in the \emph{emmeans} package in R~\cite{Searle1980PMM}. The $p$-values were adjusted using the Holm correction~\cite{Holm1979}, which controls the family-wise error rate while retaining greater statistical power than more conservative approaches. All mixed-effects models demonstrated satisfactory convergence~\cite{Bates2015lme4}, with no singular fits, no convergence warnings, and maximum absolute optimization gradients below 0.001.

\begin{figure*}[htbp]
  \centering

  % ---------- Row 1 ----------
  \begin{subfigure}[b]{0.33\linewidth}
    \centering
    \includegraphics[width=\linewidth]{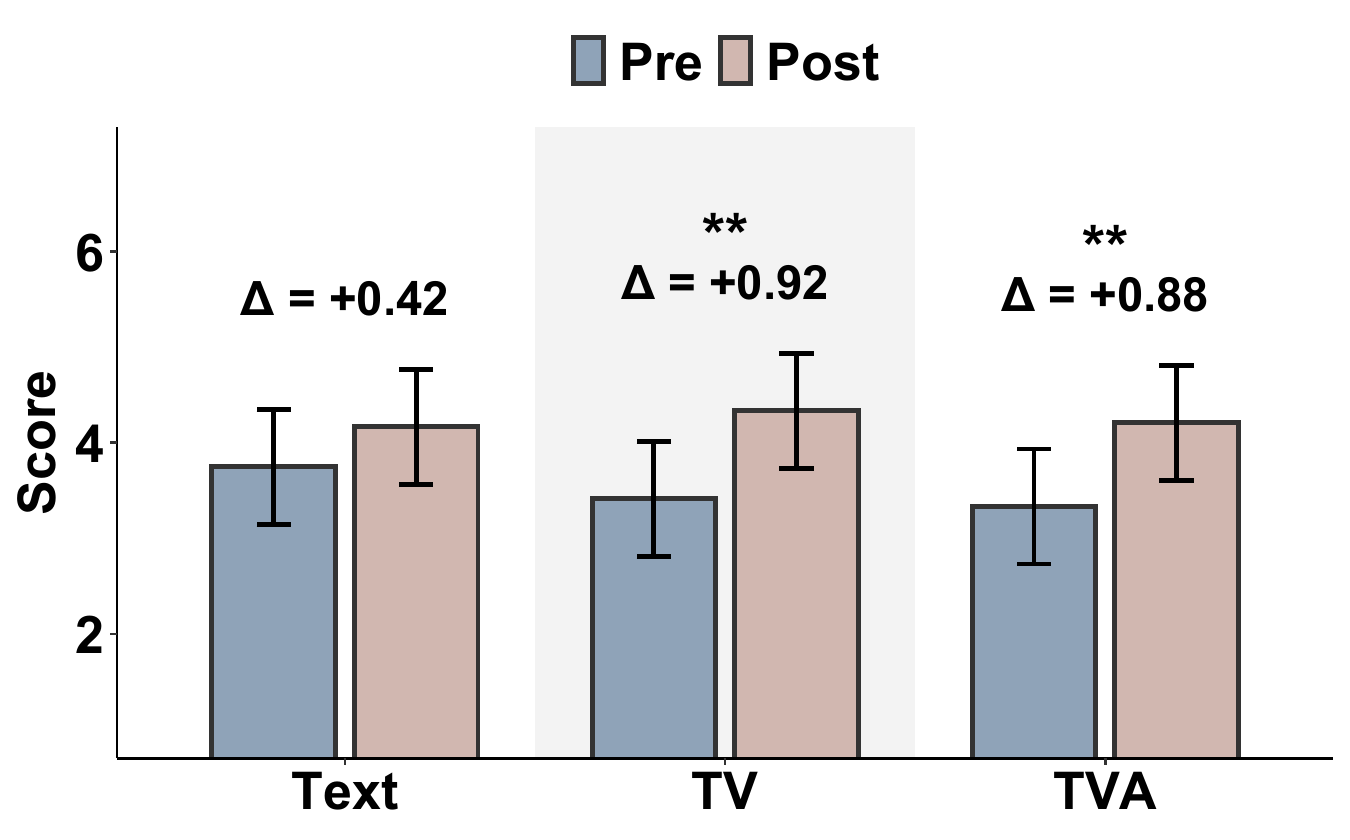}
    \caption{Arousal}
    \label{fig:arousal_short}
  \end{subfigure}
  \hfill
  \begin{subfigure}[b]{0.33\linewidth}
    \centering
    \includegraphics[width=\linewidth]{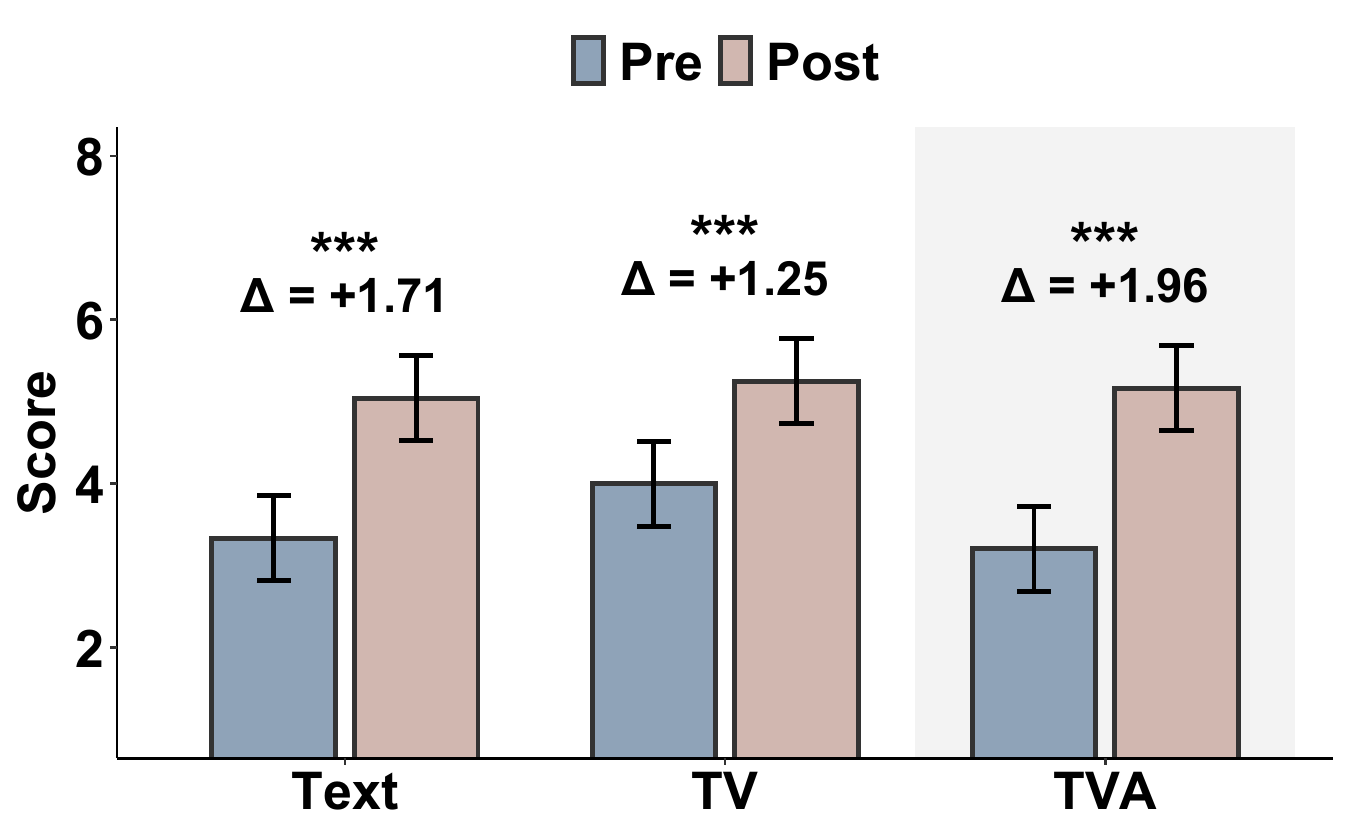}
    \caption{Valence}
    \label{fig:valence_short}
  \end{subfigure}
  \hfill
  \begin{subfigure}[b]{0.33\linewidth}
    \centering
    \includegraphics[width=\linewidth]{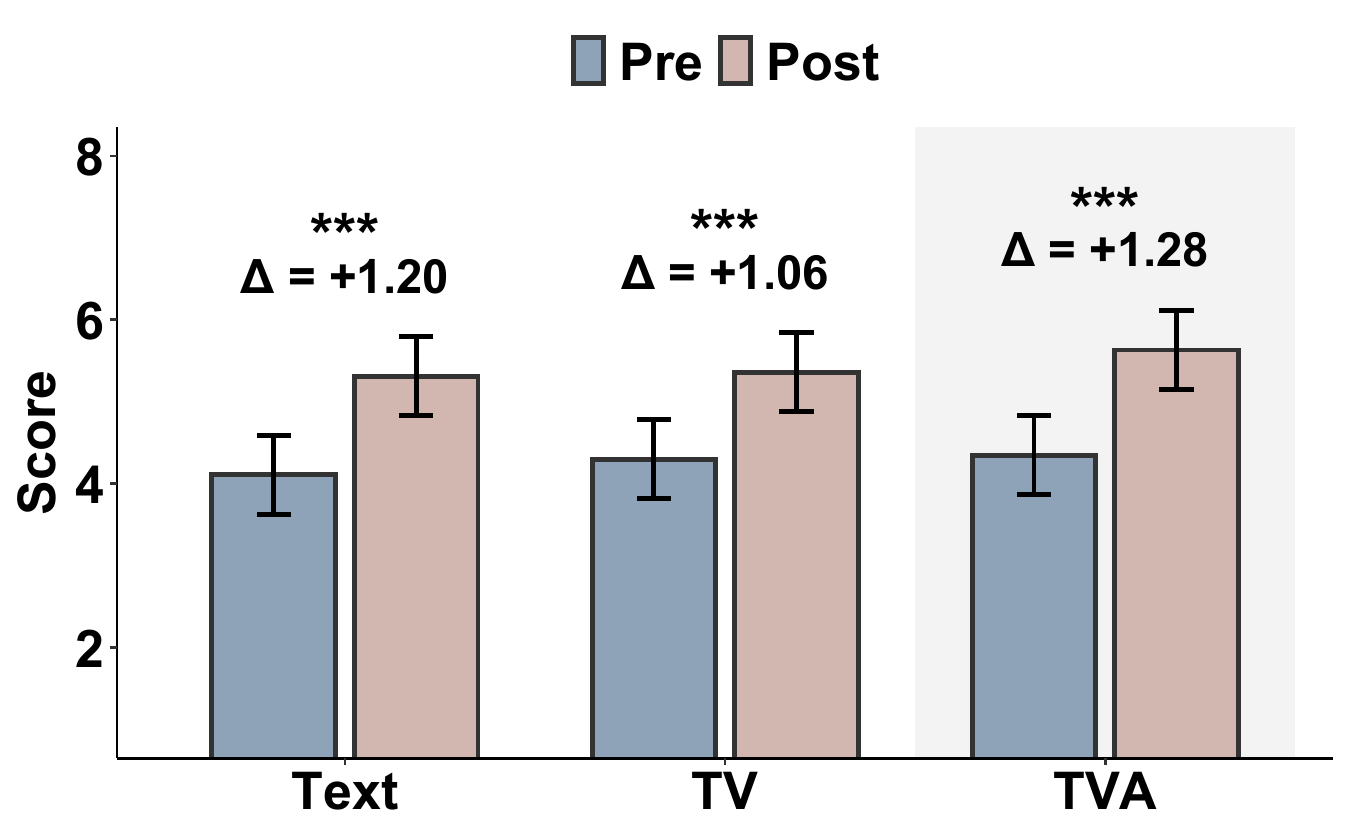}
    \caption{SHS}
    \label{fig:shs_short}
  \end{subfigure}

  \vspace{0.8em}

  % ---------- Row 2 ----------
  \begin{subfigure}[b]{0.33\linewidth}
    \centering
    \includegraphics[width=\linewidth]{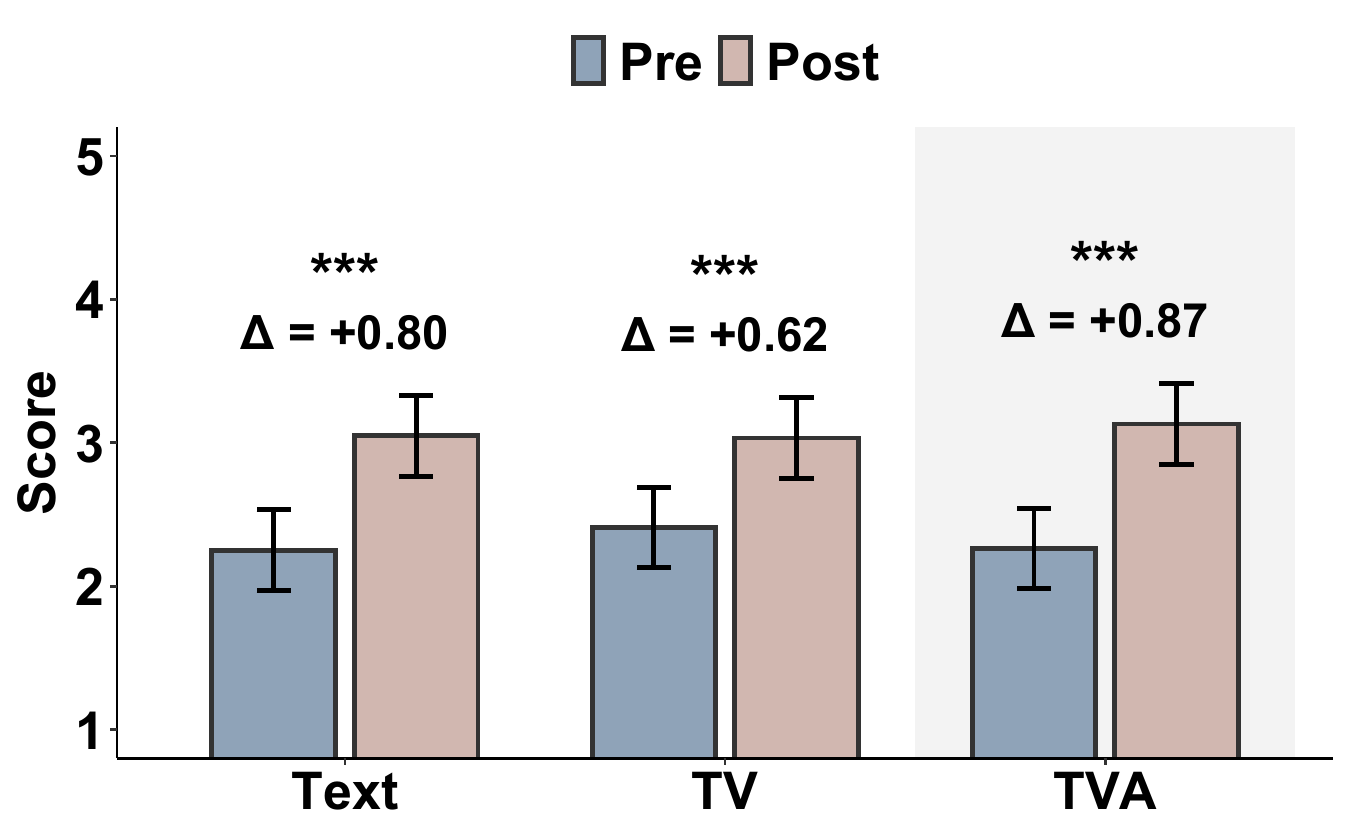}
    \caption{STAI}
    \label{fig:stai_short}
  \end{subfigure}
  \hfill
  \begin{subfigure}[b]{0.33\linewidth}
    \centering
    \includegraphics[width=\linewidth]{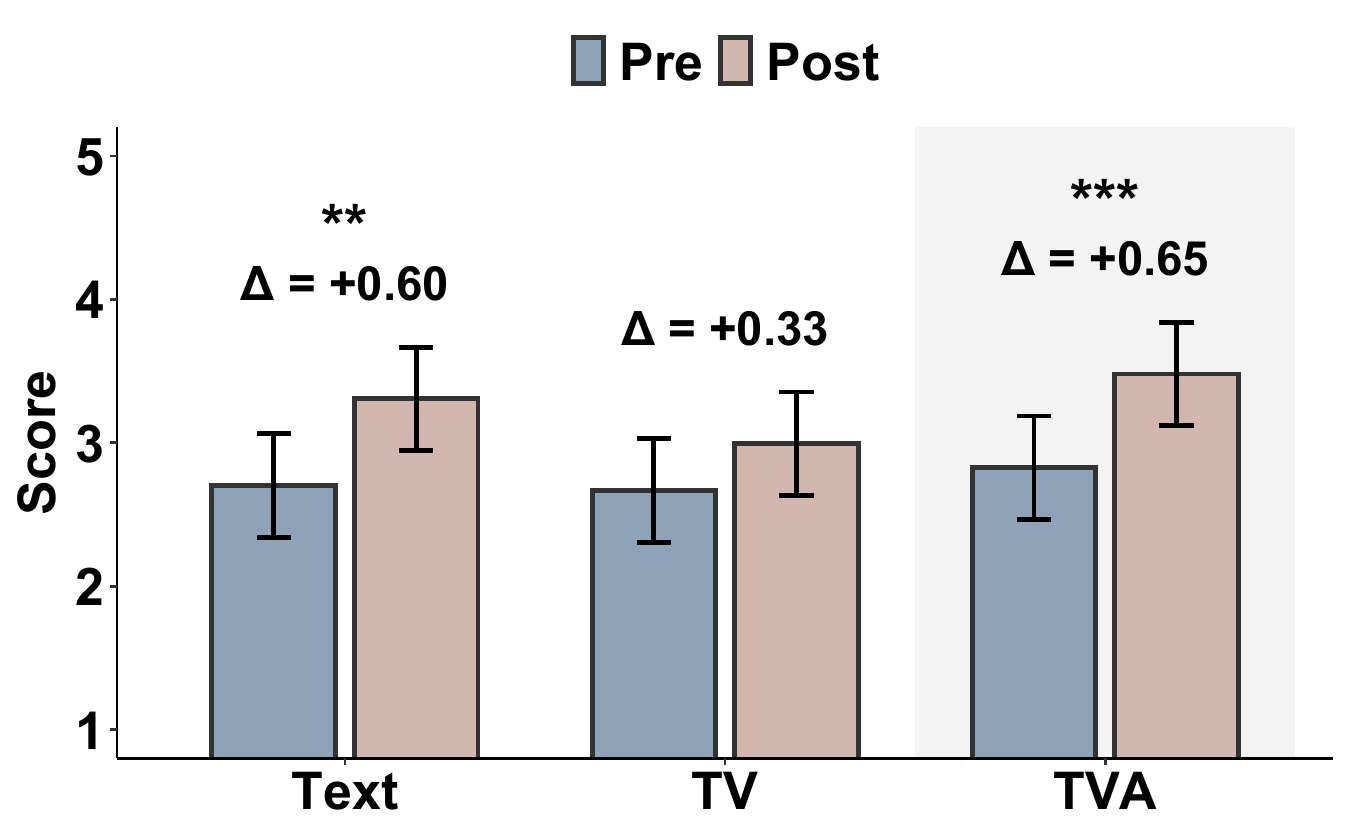}
    \caption{BRS}
    \label{fig:brs_short}
  \end{subfigure}
  \hfill
  \begin{subfigure}[b]{0.33\linewidth}
    \centering
    \includegraphics[width=\linewidth]{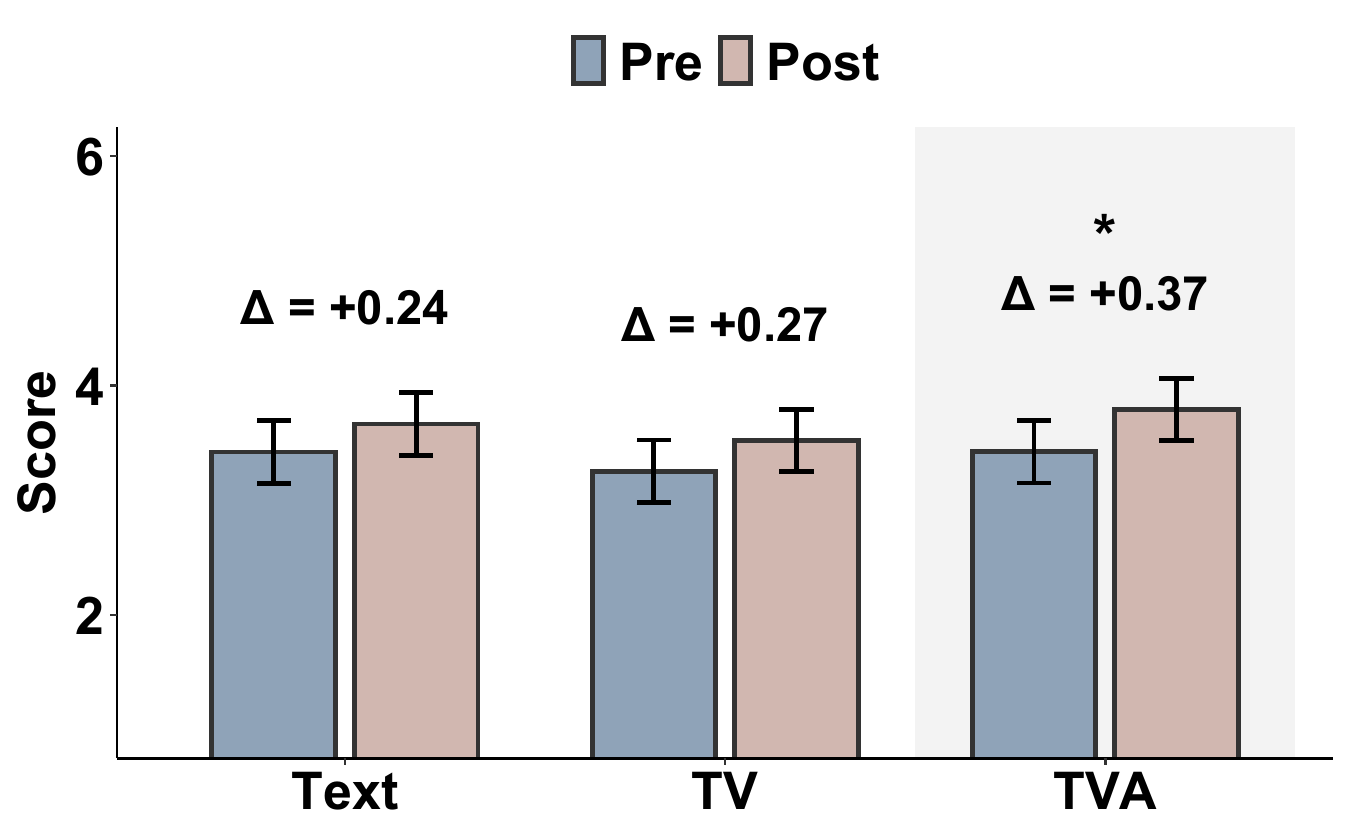}
    \caption{SSQ}
    \label{fig:ssq_short}
  \end{subfigure}

  \caption{Scenario 1: subjective questionnaire results of \emph{self-counseling} comparing pre- and post-intervention scores. Error bars: $\pm$95\% confidence intervals. $\Delta$: within-condition improvements from pre- to post-intervention (Post$-$Pre). Shaded backgrounds highlight the condition exhibiting the largest improvement. $\star$: $<0.05$, ${\star}{\star}$: $<0.01$,${\star}{\star}{\star}$: $<0.001$.}
  \label{fig:prepost_emm}
\end{figure*}

\subsection{Analysis Methods}

A within-subject design was used for the short-term study, and the same analytical framework was applied to the long-term datasets. Primary analyses employed mixed-effects models implemented with the \texttt{lme4} package in R~\cite{Bates2015lme4}. Random-intercept models were used to account for stable individual differences across repeated measurements. The independent variables included \textbf{condition} (three levels: \textbf{Text}, \textbf{Text+Voice}, and \textbf{Text+Voice+Appearance}) and \textbf{time} (two levels: pre- and post-experiment).

Post-hoc pairwise comparisons were conducted using estimated marginal means via the \emph{emmeans} package in R~\cite{Searle1980PMM}. Holm correction was applied to adjust $p$-values and control the family-wise error rate while maintaining statistical power~\cite{Holm1979}. All models converged successfully, with no singular fits or convergence warnings and optimization gradients below 0.001.

% \section{Results}

% Results from the short-term \emph{self-counseling} study and the long-term \emph{self-compassion} study are reported and analyzed using two perspectives: \textbf{subjective} and \textbf{objective} assessments. \textbf{Subjective assessments} are based on participants’ responses to self-report questionnaires (Section~\ref{sec:Procedure}), which capture perceived emotional states and psychological changes from the participant’s perspective. As self-report data may be influenced by response biases and limitations of introspection, these measures are interpreted in conjunction with objective indicators rather than in isolation.
% \textbf{Objective assessments} complement self-report by using measures that are less dependent on participants’ self-perception. Specifically, behavioral metrics derived from the chat interaction (e.g., total dialogue duration and conversational turns) and physiological indicators derived from ECG signals were analyzed. Collectively, these measures provide additional evidence regarding engagement, behavioral patterns, and physiological regulation.

\section{Results}

Results from the short-term \emph{self-counseling} and long-term \emph{self-compassion} studies are analyzed from two perspectives: \textbf{subjective} and \textbf{objective}.  
\textbf{Subjective assessments} are derived from participants’ questionnaire responses (Section~\ref{sec:Procedure}), capturing perceived emotional and psychological states. Because self-reports may be influenced by response bias and introspection limits, they are interpreted alongside objective indicators.
% \textbf{Objective assessments} complement self-reports using behavioral and physiological measures. Behavioral metrics from chat interactions (e.g., dialogue duration and conversational turns) reflect engagement, while physiological indicators derived from ECG signals provide evidence of autonomic regulation. Together, these measures offer a more comprehensive evaluation of user experience and emotional outcomes.
\textbf{Objective assessments} complement self-reports using behavioral. Behavioral metrics from chat interactions (e.g., dialogue duration and conversational turns) reflect engagement.

\subsection{Results Analysis on Short-term Self-counseling Intervention}

\subsubsection{\bf Subjective Questionnaire Results}

Fig.~\ref{fig:prepost_emm} presents the subjective questionnaire results from the short-term \emph{self-counseling} study by comparing pre- and post-intervention scores across three experimental conditions: \textbf{Text}, \textbf{Text+Voice (TV)}, and \textbf{Text+Voice+Appearance (TVA)}. All questionnaire items were scored following standard procedures, including reverse-scoring of negatively worded items (e.g., for 7-point scales, responses were transformed using $8-x$ where applicable). Higher scores consistently indicate more desirable outcomes across all measures shown in the figure (e.g., more positive affect, reduced anxiety, enhanced resilience, or decreased discomfort).

\paragraph{\bf Overall trends across conditions.}
As shown by the significance markers in Fig.~\ref{fig:prepost_emm}, most subjective measures exhibited statistically significant pre--post differences. Many effects reached $p<0.01$ or $p<0.001$ (denoted by $\star$, $\star\star$, and ${\star}{\star}{\star}$), indicating consistent improvements following the \emph{self-counseling} intervention.
Specifically, post-intervention scores increased in \emph{Valence} and \emph{Arousal}, reflecting more positive emotional states and stronger engagement. Scores on \emph{SHS} and \emph{BRS} also increased, suggesting enhanced motivation, clearer coping pathways, and greater perceived resilience. Because \emph{STAI} was reverse-scored, higher post-intervention scores indicate reduced anxiety. Similarly, reverse-coded \emph{SSQ} results show that simulator sickness did not increase and was reduced for participants who initially experienced discomfort. Overall, these consistent changes across measures indicate that the intervention effectively supported short-term emotional regulation and coping.

\paragraph{\bf Comparison among different conditions.}
Although all conditions showed positive pre--post changes, the magnitude varied, with richer self-representation generally producing stronger effects.
For \emph{Valence} and \emph{Arousal} (Fig.~\ref{fig:valence_short}, Fig.~\ref{fig:arousal_short}), all conditions exhibited significant increases. \textbf{TVA} produced the largest improvement in \emph{Valence}, suggesting that interacting with a visually embodied self-resembling avatar most strongly promoted positive emotional appraisal. In contrast, \textbf{TV} produced the largest increase in \emph{Arousal}, consistent with evidence that self-similar voice cues enhance immediacy and social salience~\cite{Huang2023EmotionalSelfVoice}. The presence of a highly familiar visual self-representation may instead increase comfort and reduce arousal~\cite{Coan2006SocialRegulation}.
For \emph{SHS} (Fig.~\ref{fig:shs_short}), the largest improvement occurred in the \textbf{TVA} condition, indicating stronger gains in perceived agency and coping pathways when both visual and vocal self-cues were present. A similar pattern appeared for \emph{STAI} (Fig.~\ref{fig:stai_short}), where \textbf{TVA} produced the greatest reduction in anxiety.
For \emph{BRS} (Fig.~\ref{fig:brs_short}), participants in the \textbf{TVA} condition also reported the largest increase in perceived resilience, suggesting that richer embodiment may enhance users’ perceived ability to recover from stress.
Finally, for \emph{SSQ} (Fig.~\ref{fig:ssq_short}), \textbf{TVA} showed the greatest improvement, indicating that visual embodiment did not introduce additional discomfort and may instead enhance comfort and stability during VR interaction.

\paragraph{\bf Interaction experience assessment.}
Fig.~\ref{fig:NM-SP} compares perceived social presence across conditions using the \emph{NM-SP}. Scores were highest in the \textbf{TVA} condition, followed by \textbf{TV}, with \textbf{Text} showing the lowest values. This trend indicates that richer self-representation increased the perceived authenticity and social presence of the interaction.
The \textbf{TVA} condition exhibited the highest model-estimated mean and a relatively concentrated distribution, suggesting that the visually embodied self-avatar enhanced co-presence and interpersonal realism. In contrast, the \textbf{Text} condition showed lower central tendency and greater variability, indicating less consistent perceptions of social presence when interaction relied solely on textual dialogue.

% \subsubsection{\bf Objective Results}
% \paragraph{\bf Intervention engagement assessment.}
% Fig.~\ref{fig:chat} presents key statistics derived from analyzing the conversations between participants and their \mirror{}s during \emph{self-counseling}, specifically chat duration (in seconds) and the number of conversational turns. These metrics serve as objective behavioral indicators of participants' engagement throughout the \emph{self-counseling} intervention. Unlike self-report measures, they quantify the duration and frequency of participants' interaction with the system, thereby reflecting sustained involvement without relying on subjective appraisal.

\subsubsection{\bf Objective Results}
\begin{figure}[htb]
  \centering

 \begin{subfigure}[b]{0.45\linewidth}
    \centering
    \includegraphics[width=\linewidth]{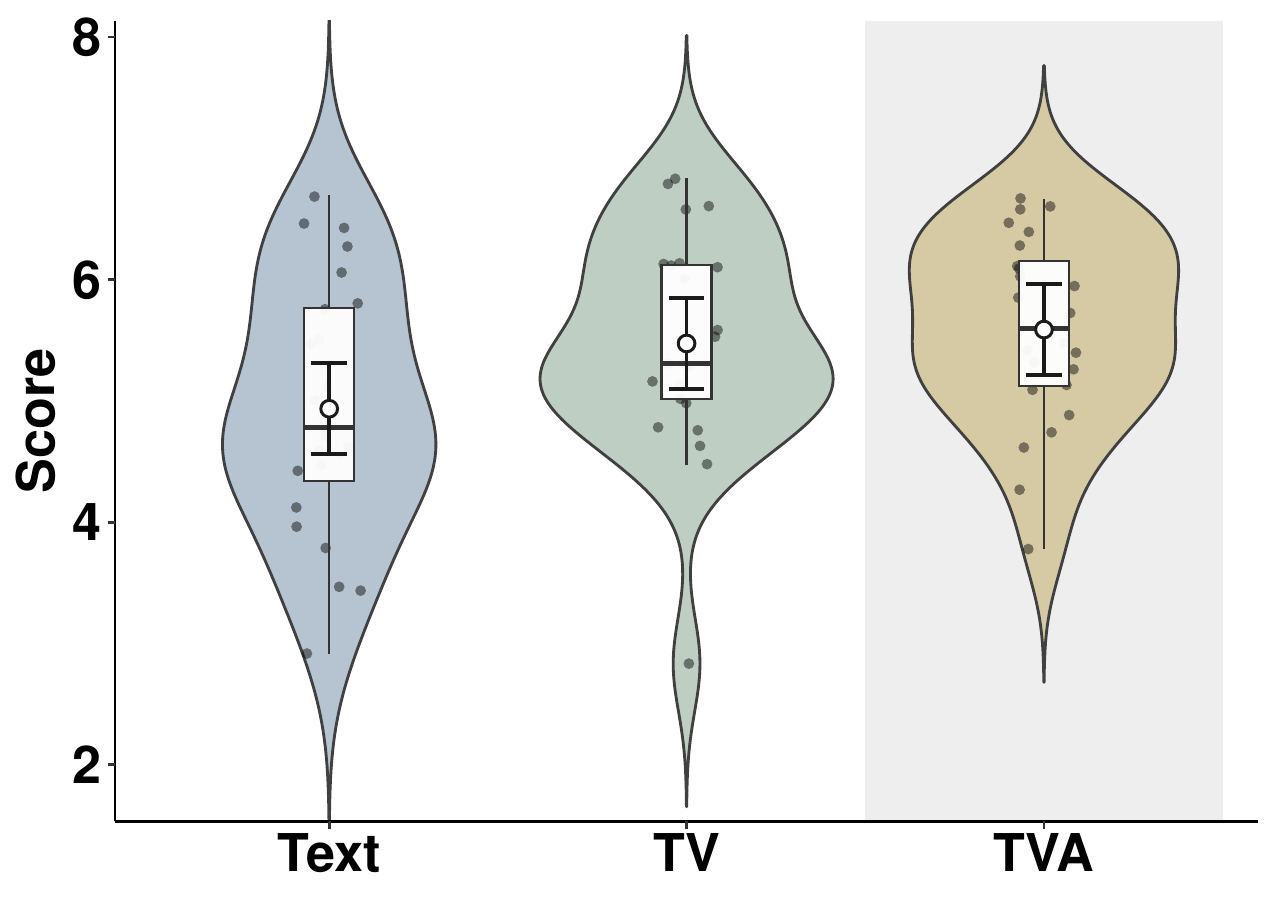}
    \caption{Network minds measures}
    \label{fig:NM-SP}
  \end{subfigure}

  \begin{subfigure}[b]{0.45\linewidth}
    \centering
    \includegraphics[width=1\linewidth]{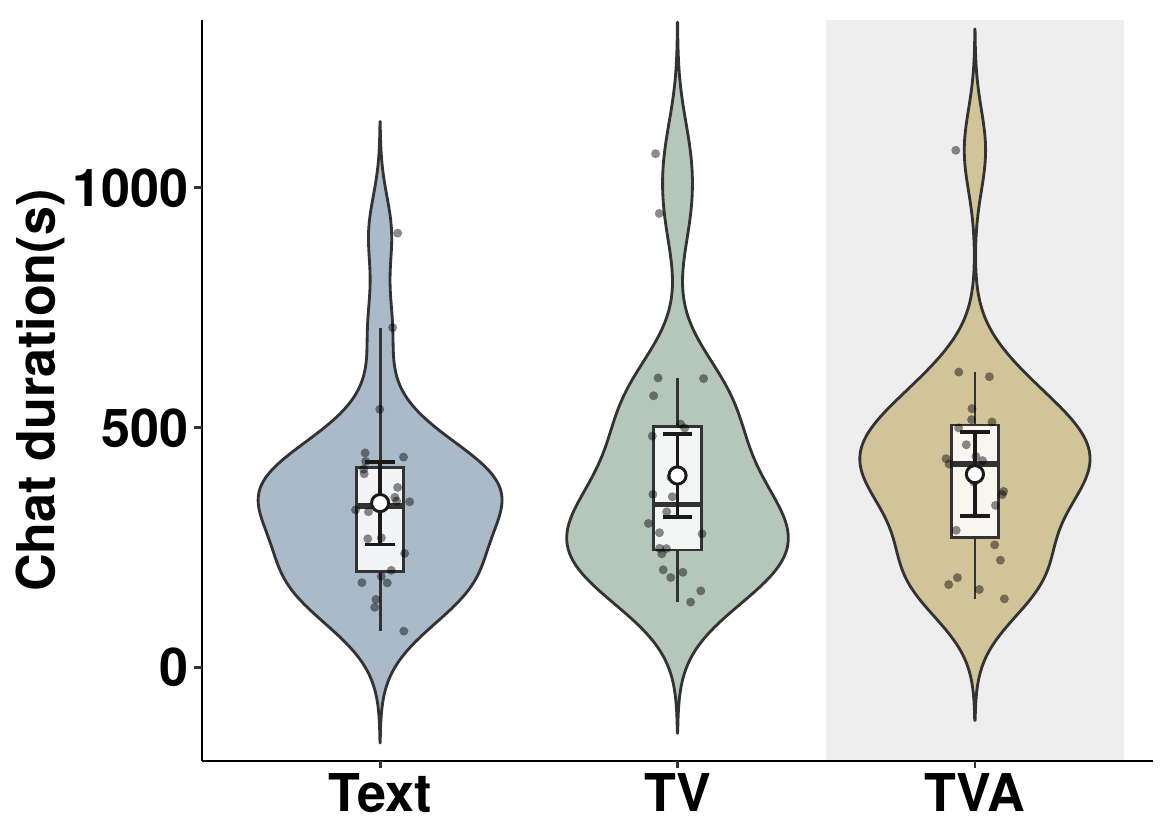}
    \caption{Chat duration}
    \label{fig:chat duration}
  \end{subfigure}
  \begin{subfigure}[b]{0.45\linewidth}
    \centering
    \includegraphics[width=1\linewidth]{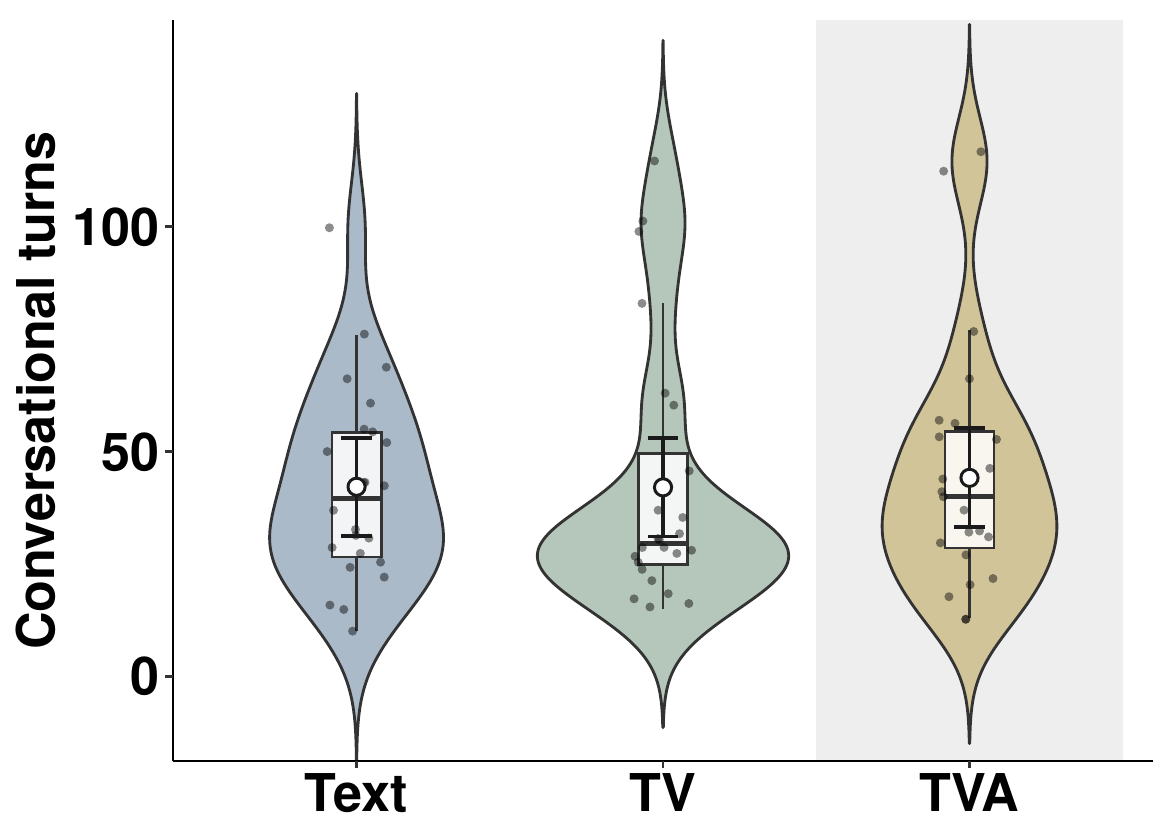}
    \caption{Conversational turns}
    \label{fig:utterance count}
  \end{subfigure}

  \caption{
  Scenario 1: (a) Networked Minds Measure of Social Presence (NM-SP) scores across experimental conditions. Error bars: $\pm$95\% confidence intervals. Boxplots: median and interquartile range. Shaded backgrounds highlight the condition with the highest estimated social presence. (b) Chat duration and (c) conversational turns across different experimental conditions.
  Error bars indicate $\pm$95\% confidence intervals. Boxplots show the median and interquartile range. 
  Shaded backgrounds highlight the condition with the highest estimated conversational engagement.
  }
  \label{fig:chat}
\end{figure}

\paragraph{\bf Intervention engagement assessment.}
Fig.~\ref{fig:chat duration} and Fig.~\ref{fig:utterance count} summarizes conversation statistics during the \emph{self-counseling} intervention, including chat duration (seconds) and the number of conversational turns. These metrics provide objective indicators of participant engagement by quantifying the length and frequency of interactions with the \mirror{}, complementing subjective assessments without relying on self-report.

% Across all conditions, participants generally engaged in extended conversations, indicating that the system successfully supported multi-turn dialogue rather than brief, one-shot exchanges. Notably, both engagement metrics revealed systematic differences between experimental conditions. As shown in Fig.~\ref{fig:chat duration}, the \textbf{TVA} condition showed the longest chat durations on average, with higher model-estimated means and a more concentrated distribution compared to the \textbf{Text} and \textbf{TV} conditions. These findings indicate that interaction with a visually embodied, self-resembling \mirror{} encouraged participants to remain in the conversation for longer periods.

% As shown in Fig.~\ref{fig:utterance count}, a similar pattern emerged for the number of conversational turns. Although all conditions elicited active participation, the \textbf{TVA} condition again exhibited the highest model-estimated number of conversational turns, indicating more frequent reciprocal exchanges. This pattern suggests that richer social cues, such as voice and visual embodiment, enhance mutual understanding and increase social salience, thereby making the interaction feel more like a genuine dialogue rather than a purely cognitive task.

Across conditions, participants generally engaged in extended conversations, indicating that the system supported sustained multi-turn dialogue. Both engagement metrics showed systematic differences across conditions. As shown in Fig.~\ref{fig:chat duration}, the \textbf{TVA} condition produced the longest chat durations, with higher model-estimated means and a more concentrated distribution than the \textbf{Text} and \textbf{TV} conditions. This suggests that interacting with a visually embodied, self-resembling \mirror{} encouraged longer engagement.
A similar pattern was observed for conversational turns (Fig.~\ref{fig:utterance count}). While all conditions elicited active participation, the \textbf{TVA} condition showed the highest number of turns, indicating more frequent exchanges. These results suggest that richer social cues, including voice and visual embodiment, increased social salience and promoted more interactive dialogue.

% \paragraph{\bf ECG-based assessment.}

% \begin{figure}[htb]
% \centering

% % ---------- First Row ----------
% \begin{subfigure}[b]{0.5\linewidth}
%   \centering
%   \includegraphics[width=\linewidth]{figures/hr_mean.pdf}
%   \caption{HR mean}
%   \label{fig:HR_mean}
% \end{subfigure}

% \vspace{0.6em}

% % ---------- Second Row ----------
% \begin{subfigure}[b]{0.49\linewidth}
%   \centering
%   \includegraphics[width=\linewidth]{figures/sdnn.pdf}
%   \caption{SDNN}
%   \label{fig:SDNN}
% \end{subfigure}
% % \hfill
% \begin{subfigure}[b]{0.49\linewidth}
%   \centering
%   \includegraphics[width=\linewidth]{figures/PNN50.pdf}
%   \caption{PNN50}
%   \label{fig:PNN50}
% \end{subfigure}

% \caption{Scenario 1: post-interaction changes in HR and HRV metrics under different conditions. Error bars indicate $\pm$95\% confidence intervals. $\Delta$: Post$-$Pre. Shaded backgrounds highlight the condition exhibiting the largest improvement.}
% \label{fig:ecg_after-before}
% \end{figure}
\begin{figure*}[htbp]
  \centering

  % 整体缩进，控制左右留白
  % \begin{minipage}{1\linewidth}
    \centering

    % ================== Row 1: 2 figures (centered) ==================
    % \hspace{0.005\linewidth} % 左侧空一列（0.30 + 0.035 ≈ 0.335 / 2）
    \begin{subfigure}[b]{0.33\linewidth}
      \centering
      \includegraphics[width=\linewidth]{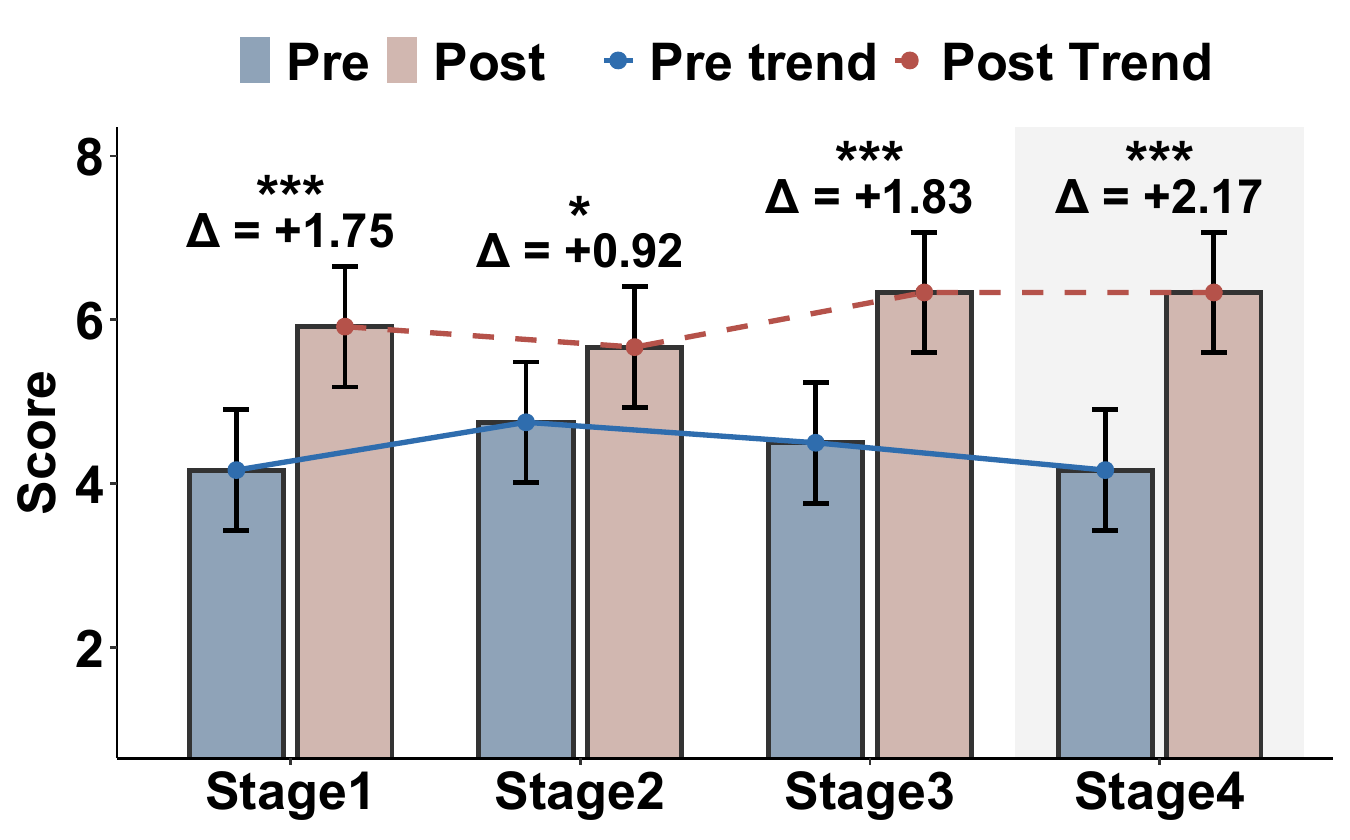}
      \caption{Valence}
      \label{fig:valence}
    \end{subfigure}
    % \hspace{0.035\linewidth}
    \begin{subfigure}[b]{0.33\linewidth}
      \centering
      \includegraphics[width=\linewidth]{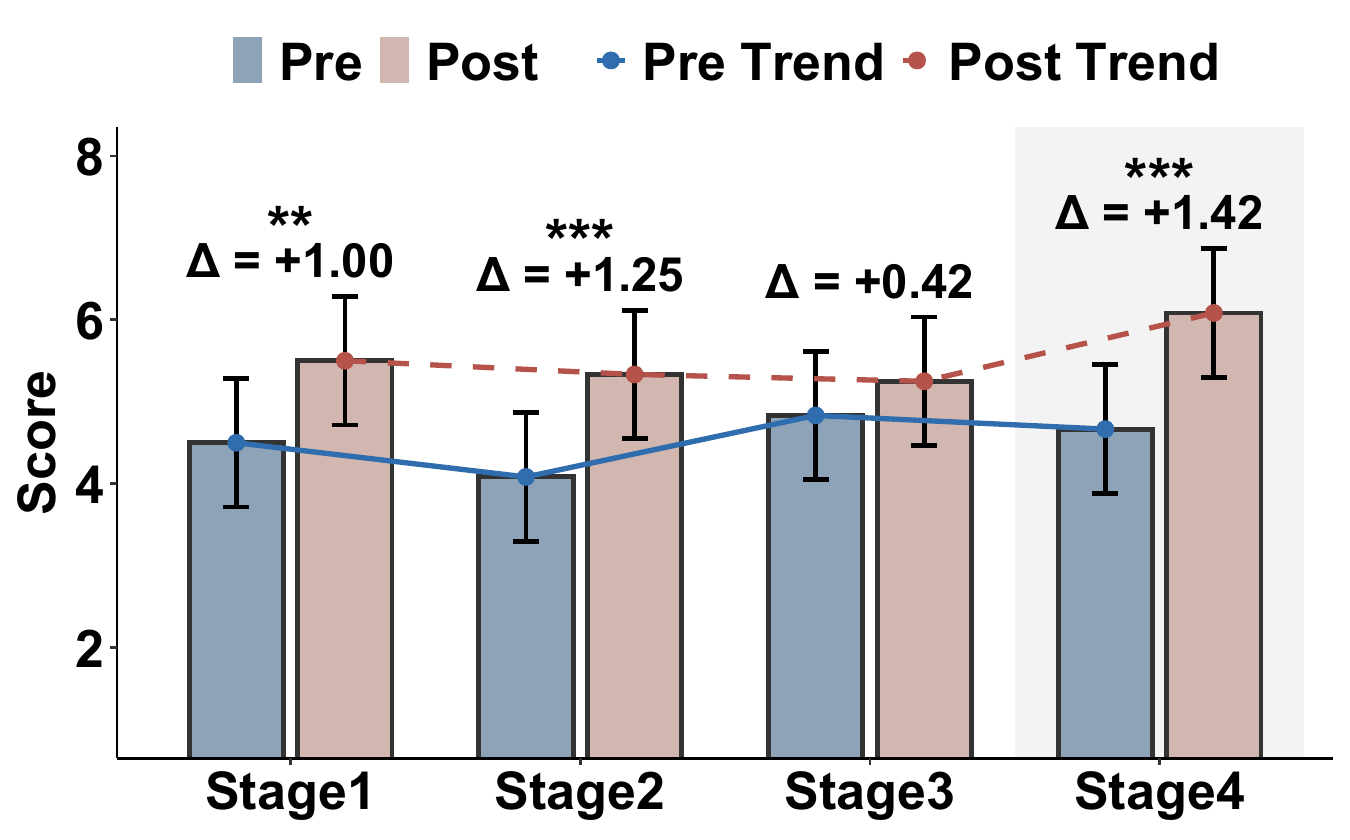}
      \caption{Arousal}
      \label{fig:arousal}
    \end{subfigure}
    \begin{subfigure}[b]{0.33\linewidth}
      \centering
      \includegraphics[width=\linewidth]{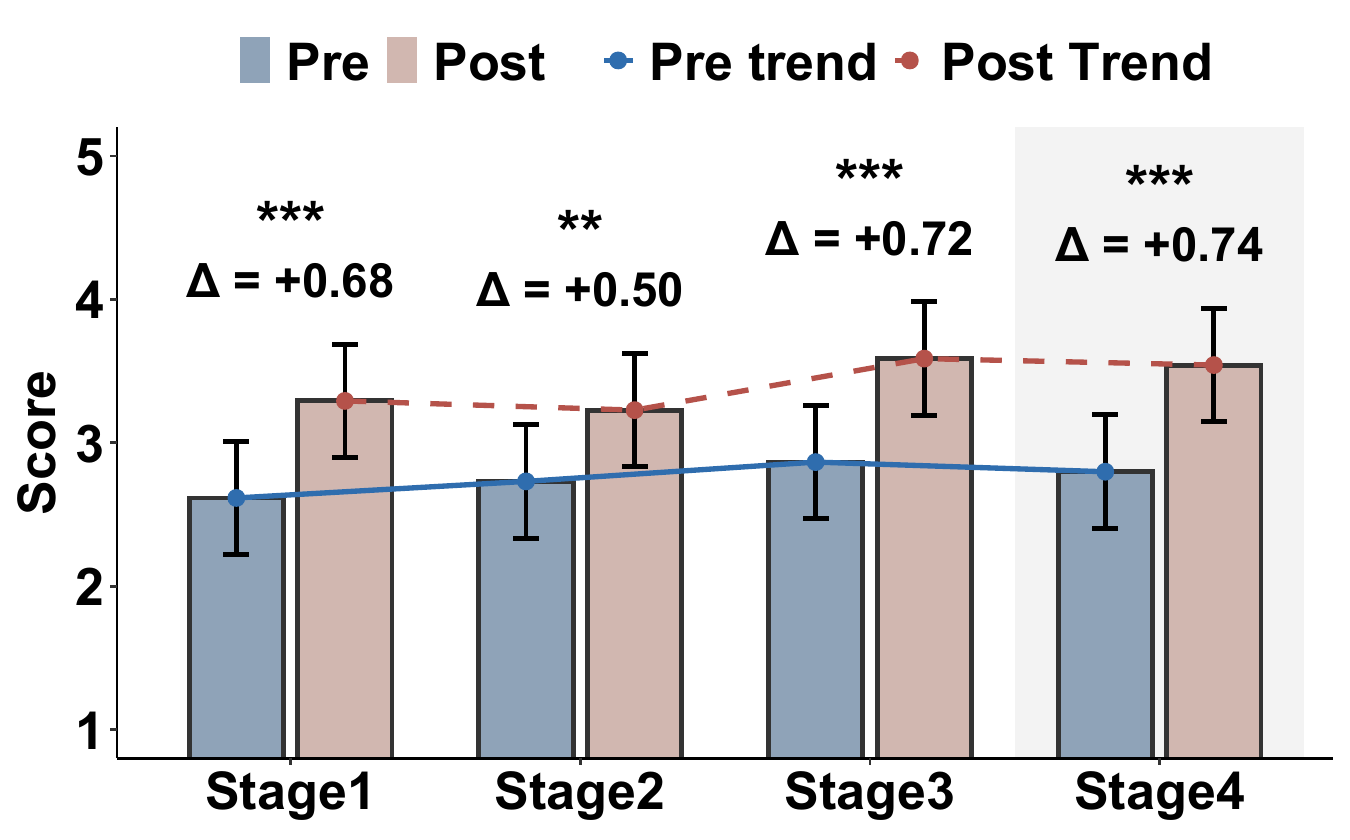}
      \caption{State SCS}
      \label{fig:state scs}
    \end{subfigure}
    % \hspace{0.0\linewidth}
    \begin{subfigure}[b]{0.33\linewidth}
      \centering
      \includegraphics[width=\linewidth]{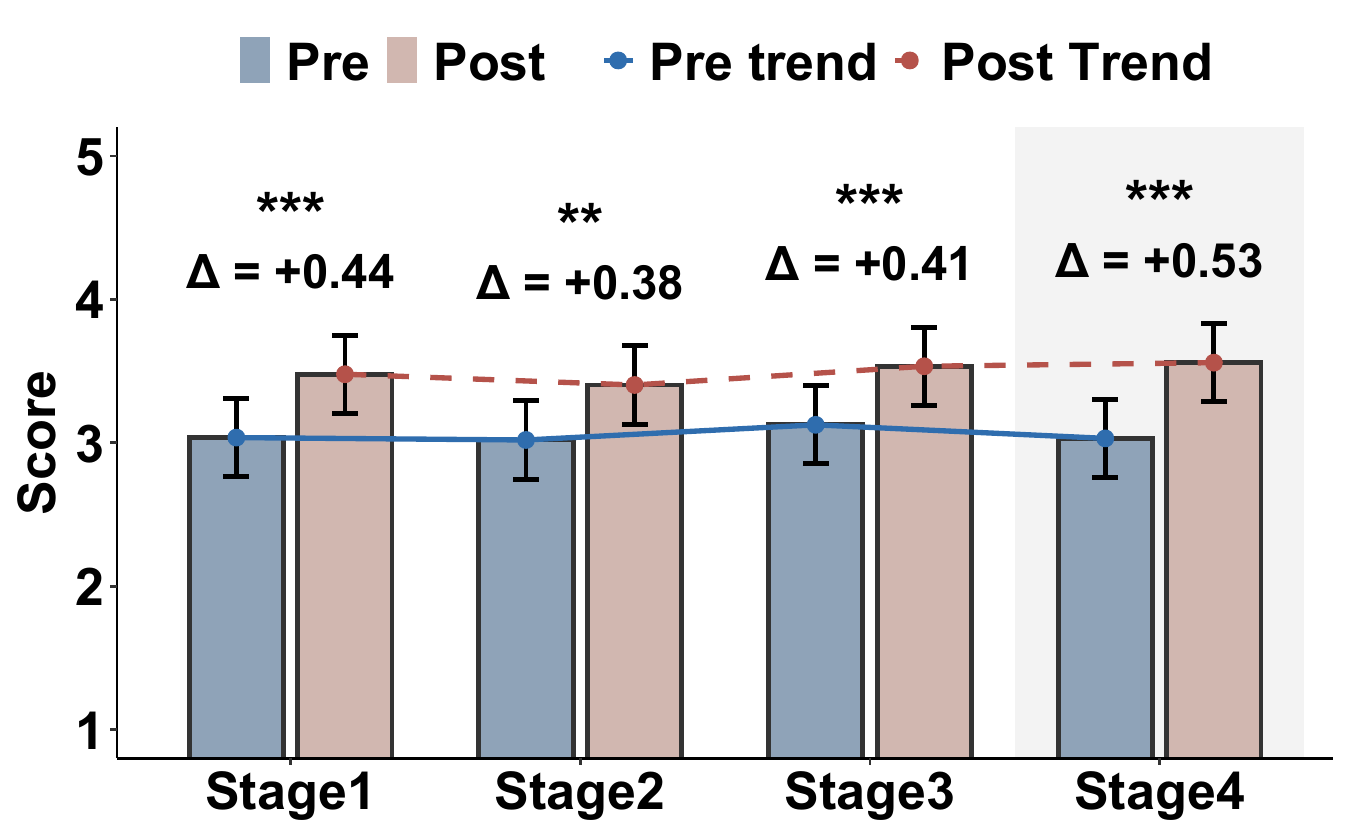}
      \caption{NMR}
      \label{fig:nmr}
    \end{subfigure}
    % \hspace{0.035\linewidth}
    \begin{subfigure}[b]{0.33\linewidth}
      \centering
      \includegraphics[width=\linewidth]{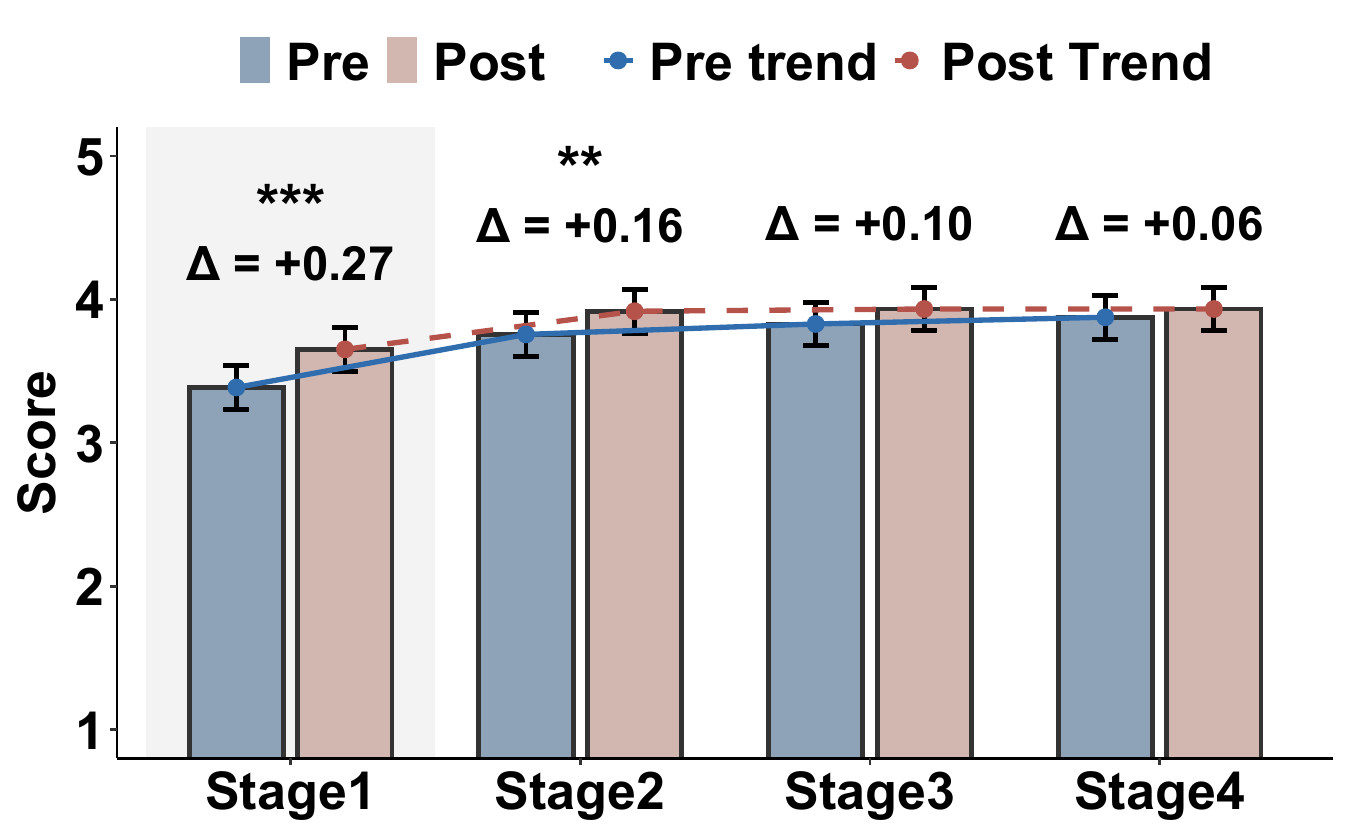}
      \caption{SSQ}
      \label{fig:ssq}
    \end{subfigure}
    \begin{subfigure}[b]{0.33\linewidth}
      \centering
      \includegraphics[width=\linewidth]{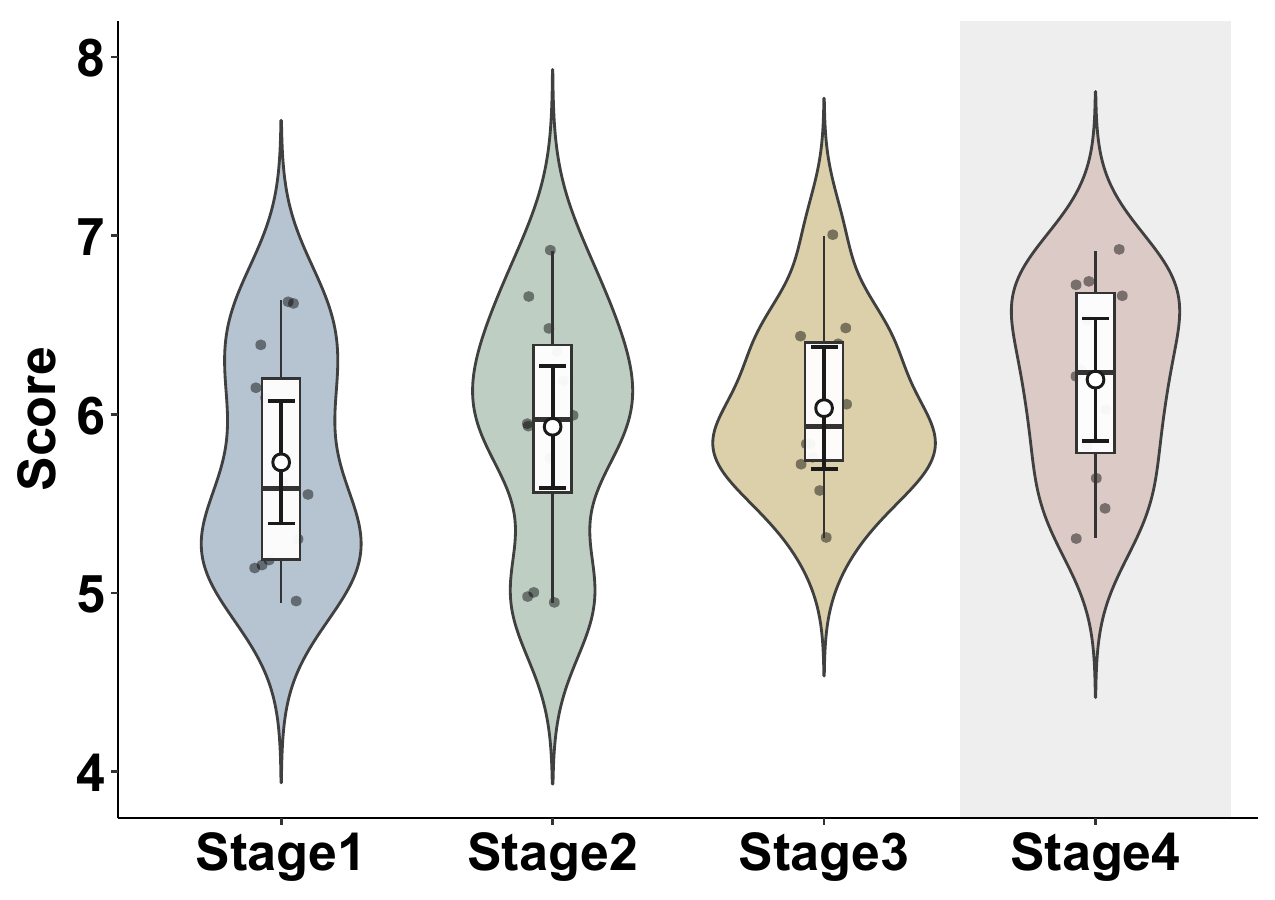}
      \caption{NM-SQ}
      \label{fig:NM-SP_longterm}
    \end{subfigure}
  % \end{minipage}

  \caption{Scenario 2: (a)-(e) subjective questionnaire results of \emph{self-compassion} comparing pre- and post-intervention scores. Error bars: $\pm$95\% confidence intervals. $\Delta$: within-condition improvements from pre- to post-intervention (Post$-$Pre). Shaded backgrounds highlight the condition exhibiting the largest improvement. $\star$:$<0.05$, ${\star}{\star}$:$<0.01$,${\star}{\star}{\star}$:$<0.001$. (f) Networked Minds Measure of Social Presence (NM-SP) scores across experimental conditions. Error bars: $\pm$95\% confidence intervals. Boxplots: median and interquartile range. Shaded backgrounds highlight the condition with the highest estimated social presence.}
  \label{fig:sc2_scale}
\end{figure*}

\subsection{Results Analysis on Long-term Self-compassion Intervention}

\paragraph{\bf Overall effectiveness across stages.}
Fig.~\ref{fig:sc2_scale} shows the longitudinal questionnaire results from the \emph{self-compassion} study. Across all four stages, participants exhibited consistent pre--post improvements on most measures, indicating reliable short-term benefits in each session. Most session-level gains reached statistical significance (often $p<0.01$ or $p<0.001$). The final session (Stage~4) showed the strongest effects, with all measures except SSQ demonstrating highly significant improvements (${\star}{\star}{\star}$), suggesting that repeated exposure maintained and potentially strengthened the intervention’s effectiveness.

\paragraph{\bf Stage-wise patterns and cumulative trends.}
Beyond within-session effects, the dashed trend lines indicate a clear longitudinal pattern: post-intervention scores generally increased with stage for key outcomes (Valence, Arousal, State SCS, and NMR). This upward trend suggests cumulative benefits from repeated weekly sessions. In contrast, pre-session scores varied across stages and should be interpreted cautiously, as the week-long intervals introduced uncontrolled external influences that may affect baseline emotional states.

\paragraph{\bf Measure-specific interpretations.}
For \emph{Valence} and \emph{Arousal} (Fig.~\ref{fig:valence}, Fig.~\ref{fig:arousal}), significant pre--post increases were observed at every stage, indicating consistent improvements in affective experience. The increasing post-session trajectory across stages suggests that repeated interactions progressively facilitated more positive appraisal and emotional engagement.
For \emph{State SCS} (Fig.~\ref{fig:state scs}) and \emph{NMR} (Fig.~\ref{fig:nmr}), all sessions showed positive gains, with larger improvements emerging in later stages. The upward post-session trend suggests that repeated use strengthened participants’ self-compassion and perceived ability to regulate negative emotions.
For \emph{SSQ} (Fig.~\ref{fig:ssq}), improvements were more pronounced in earlier stages and diminished later. This pattern likely reflects a ceiling effect, as pre-session SSQ scores in later stages were already high.

\paragraph{\bf Interaction experience assessment.}
Fig.~\ref{fig:NM-SP_longterm} shows perceived social presence (\emph{NM-SP}) across the four weekly sessions. Scores remained consistently high, indicating that participants generally experienced the interaction as socially present and authentic. Model-estimated summaries also reveal a gradual increase across stages, with the highest social presence observed at Stage~4, suggesting that repeated exposure strengthened participants’ sense of co-presence and interpersonal realism.

\subsection{System Characterizations}

The quality of the reconstructed appearance and cloned voice of the \mirror{} in \systemname{} was accessed to evaluate users’ subjective satisfaction with the system. Participants rated both the \mirror’s appearance and the cloned voice prior to the intervention. All ratings were collected using a 5-point Likert scale (1 = very low, 5 = very high) across four dimensions: \emph{Familiarity}, \emph{Similarity}, \emph{Naturalness}, and overall \emph{Quality}. The results are summarized in Table~\ref{tab:rating}.

\begin{table}[htb]
  \caption{Participants’ ratings of the appearance and cloned voice of their \mirror{} across four perceptual dimensions on a 5-point Likert scale, where a higher score indicates better evaluation.}
  \label{tab:rating}
  \scriptsize%
  \centering%
  \begin{tabu}{
    l
    *{2}{c}
    *{2}{c}
  }
  \toprule
   & \multicolumn{2}{c}{\textbf{Appearance}} 
   & \multicolumn{2}{c}{\textbf{Voice}} \\
   & Mean & SD & Mean & SD \\
  \midrule
  \textbf{Familiarity}  & 3.50 & 1.07 & 4.08 & 0.89 \\
  \textbf{Similarity}   & 3.42 & 1.04 & 4.08 & 0.89 \\
  \textbf{Naturalness}  & 3.64 & 0.95 & 4.36 & 0.75 \\
  \textbf{Quality}      & 3.92 & 0.79 & 4.44 & 0.64 \\
  \bottomrule
  \end{tabu}%
\end{table}

\paragraph{\bf Appearance reconstruction quality.}
Participants first evaluated the appearance of their reconstructed virtual avatar by viewing it prior to the user study. As shown in Table~\ref{tab:rating}, appearance-related ratings were generally high across all four dimensions. Mean scores were consistently close to or above 3.5, including \emph{Familiarity} ($M=3.50$, $SD=1.07$), \emph{Similarity} ($M=3.42$, $SD=1.04$), \emph{Naturalness} ($M=3.64$, $SD=0.95$), and overall \emph{Quality} ($M=3.92$, $SD=0.79$). These results indicate that participants, on average, perceived the reconstructed avatars as recognizable, sufficiently similar to their own appearance, and appropriate for interactive use.

\paragraph{\bf Voice cloning quality.}
Participants also evaluated the cloned version of their own voice by listening to sample utterances generated by the system. As shown in Table~\ref{tab:rating}, voice-related ratings were consistently high, with mean scores exceeding 4.0 across all four dimensions. Notably, ratings for \emph{Familiarity} ($M=4.08$, $SD=0.89$) and \emph{Similarity} ($M=4.08$, $SD=0.89$) indicate that the synthesized voice closely matched participants’ expectations of their own vocal identity. High ratings for \emph{Naturalness} ($M=4.36$, $SD=0.75$) and overall \emph{Quality} ($M=4.44$, $SD=0.64$) further suggest that the voice cloning module successfully captured key vocal characteristics such as timbre and prosody. Collectively, these findings demonstrate that the cloned voice was not only recognizable but also perceived as natural and high-quality, supporting its use in emotionally grounded self-dialogue.

\paragraph{\bf System responsiveness.}
In addition to enhancing perceptual quality, our system demonstrates significant improvements in real-time performance compared to previous work. While earlier methods for self-dialogue with cloned voices often required up to {\bf 30 seconds} to generate a single response~\cite{Huang2023EmotionalSelfVoice}, our system delivers spoken responses within approximately \textbf{3 seconds} after the user completes an utterance. This low-latency response is critical for maintaining conversational flow and engagement, thereby supporting multi-turn dialogue that feels interactive rather than delayed or scripted.

\section{Discussion}

The user studies provide converging evidence that embodied self-distancing through a self-resembling avatar can support emotional regulation. We interpret these findings by examining the psychological mechanisms underlying the observed effects and discussing implications and limitations of this approach.

\subsection{Effects of Self-resembling Avatar on Emotional Support}

Compared with purely cognitive or text-based strategies, VR enables immersive sensory input and embodied interaction that externalize internal experiences and make perspective shifts more perceptually concrete.

The effectiveness of self-resembling avatars may arise from several psychological mechanisms related to self-processing. First, facial appearance acts as a key cue for identity recognition~\cite{Matthew2012DiffFace}. A self-resembling avatar increases self-relevance and self-presence~\cite{KimParkLee2023SelfSimilarSocialVR}, encouraging users to perceive the interaction as self-related rather than as communication with an external agent. This stronger self-referential processing may deepen engagement with emotionally salient content~\cite{Herbert2010Selfreference}.

Second, self-resembling avatars enable a dual representation of the self. The avatar can be perceived both as ``me'' and as an externalized representation, allowing users to adopt an observer perspective while maintaining self-relevance. This balance between psychological distance and identification is critical for effective emotional regulation~\cite{Osimo2015SigmundFreud}.

Finally, combining visual similarity with a familiar self-voice further strengthens perceptual realism and trust. Multimodal self-cues can make the interaction feel more natural and conversational, reinforcing engagement and emotional processing~\cite{Guo2024Doppelganger}. Together, these mechanisms suggest that self-resembling avatars function not only as realistic representations but also as psychologically meaningful design elements that facilitate self-distancing and emotional support.

\subsection{Future Potential for Emotional and Psychological Support}

Our findings suggest broader potential for embodied, self-resembling conversational systems as tools for emotional support. Rather than replacing professional intervention, \systemname{} illustrates how immersive self-representation can facilitate psychologically grounded practices such as self-counseling, self-compassion, and reflective dialogue in an accessible and low-effort manner. By externalizing emotional experiences through an embodied self, users may engage with difficult thoughts and feelings more effectively, supporting emotional regulation through experiential interaction.

This approach could extend to additional emotional processes. Repeated embodied self-dialogue may help users reflect on emotional patterns, rehearse coping strategies, and cultivate self-compassion and resilience over time. Because interactions rely on the user’s own appearance and voice, such systems may be particularly effective for addressing self-critical or identity-related concerns where self-relevance is essential.

More broadly, combining self-resembling embodiment with generative dialogue suggests a direction for emotional technologies that emphasize user agency and self-reflection rather than continuous monitoring or external evaluation.

\subsection{Privacy and Ethics}

Self-resembling avatar reconstruction and voice cloning raise important security, privacy, and ethical considerations. In \systemname{}, facial and voice data are collected solely to generate participant-specific avatars and voices for the study and are not shared or reused beyond the experimental setting.

To minimize privacy risks, the system uses short, purpose-specific recordings rather than continuous monitoring. Raw facial and voice data are only used during the reconstruction stage and are not included in subsequent interaction analysis. Participants were informed about the data collection and intended use of generated avatars and voices to ensure transparency.

From an ethical perspective, interacting with a self-resembling avatar may increase emotional engagement due to heightened self-referential processing. Although these effects are central to the intended self-distancing mechanism, \systemname{} is designed to preserve user comfort and agency. To maintain user comfort and agency, interactions are brief, user-initiated, and designed as reflective tools rather than authoritative entities. Future work should explore responsible deployment strategies and evaluate long-term user experience as such systems move beyond controlled study environments.

\subsection{Limitations and Future Work}

The personalized \mirror{} was reconstructed from a short facial video clip to balance fidelity with practical study duration. While the resulting avatars were generally recognizable, some subtle facial idiosyncrasies may not have been fully captured. Future work will explore improved capture and reconstruction pipelines that preserve finer facial details and expressive dynamics without increasing user burden.

Personalization in \systemname{} currently relies on a user-provided character prompt. This design prioritizes low setup cost, transparency, and user control to validate the feasibility of the interaction paradigm. However, prompt-based conditioning provides only coarse and static representations of personality and communication style. Future work may investigate adaptive, data-driven personalization methods that better capture individual linguistic and emotional characteristics while maintaining usability.

Although the short-term and longitudinal studies provide complementary insights, the longitudinal deployment was intentionally limited to four weekly sessions to minimize participant burden. Consequently, the observed trends should be interpreted as preliminary indicators rather than definitive evidence of long-term effects. Future work should explore longer and more frequent deployments to better understand sustained outcomes and intervention intensity.

% Finally, the evaluation primarily relied on subjective questionnaires, with HR and HRV as the only physiological measures. Self-report measures may introduce interpretation bias, and the relatively small sample size limits generalizability. Future studies should incorporate additional objective signals (e.g., galvanic skin response or EEG) and larger, more diverse participant samples to strengthen empirical validity.

\section{Conclusion}

In summary, we presented \systemname{}, an embodied self-distancing system that integrates self-resembling avatars, cloned self-voice, and LLM-driven real-time dialogue to support emotionally grounded self-interaction in VR. In a short-term \emph{self-counseling} study, \systemname{} produced reliable pre--post improvements in affective and coping-related outcomes, with richer self-representation yielding stronger benefits. These subjective gains were further supported by converging objective evidence, including increased user engagement.
The feasibility of longitudinal use  was further demonstrated through a four-week \emph{self-compassion} study, in which benefits persisted across sessions and post-session outcomes exhibited an upward trajectory, suggesting cumulative effects. Collectively, these findings indicate that combining embodied self-representation with generative conversational AI can translate self-distancing theory into a deployable and engaging emotional support experience. \systemname{} opens new opportunities for scalable, personalized, and immersive interventions that bridge psychological theory with embodied AI systems. It also demonstrates significant and wide-ranging potential to provide comprehensive emotional and psychological support to diverse populations in a variety of challenging contexts.

\bibliographystyle{abbrv-doi}

\bibliography{template}

\end{document}